\documentclass[preprints,review,accept,moreauthors,pdftex,10pt,a4paper]{mdpi} 

\def\mnras{{\it MNRAS}}

\def\aap{{\it A\&A}}

\def\araa{{\it ARA\&A}}
\def\pasp{{\it PASP}}

\def\pasa{{\it Publ.Astron.Soc.Austral.}}

\def\apj{{\it ApJ}}
\def\apjs{{\it ApJS}}
\def\apjl{{\it ApJL}}
\def\aj{{\it The Astronomical Journal}}
\def\prd{{\it Physical Review D}}

\def\nat{{\it Nature}}

\newcommand\simlt{\lower.5ex\hbox{$\; \buildrel < \over \sim \;$}}
\newcommand\simgt{\lower.5ex\hbox{$\; \buildrel > \over \sim \;$}}

\firstpage{1} 
\pubvolume{xx}
\issuenum{1}
\articlenumber{1}
\pubyear{2018}
\copyrightyear{2018}
\externaleditor{Academic Editor: name}
\history{Received: date; Accepted: date; Published: date}

\pdfoutput=1

\Title{Radio Galaxies at VHE energies}

\Author{Frank M. Rieger $^{1}$\orcidA{}, and Amir Levinson $^{2,3}$\orcidB{}}

\AuthorNames{Frank M. Rieger and Amir Levinson}

\address{%
$^{1}$ \quad ZAH, Institut f\"ur Theoretische Astrophysik, Heidelberg University, Philosophenweg 12, 69120 Heidelberg, Germany; f.rieger@uni-heidelberg.de\\
$^{2}$ \quad The Raymond and Beverly Sackler School of Physics and Astronomy, Tel Aviv University, Tel Aviv 69978, 
                   Israel; Levinson@tauex.tau.ac.il\\
$^{3}$ \quad Yukawa Institute for Theoretical Physics, Kyoto University, Oiwake-cho, Kitashirakawa, Sakyo-ku, 
                    Kyoto 606-8502, Japan}

\abstract{Radio Galaxies have by now emerged as a new $\gamma$-ray emitting source class on the extragalactic sky. 
Given their remarkable observed characteristics, such as unusual gamma-ray spectra or ultrafast VHE variability, they 
represent unique examples to probe into the nature and physics of active galactic nuclei (AGN) in general. This review 
provides a compact summary of their observed characteristics at very high $\gamma$-ray energies (VHE; $> 100$ GeV) 
along with a discussion of their possible physics implications. A particular focus is given to a concise overview of fundamental 
concepts concerning the origin of variable VHE emission, including recent developments in black hole gap physics.} 

\begin{document}
\section{Introduction}
The current decade has seen a tremendous progress in the extragalactic Gamma-Ray Astronomy. Numerous new 
sources have been discovered by the current generation of instruments, sometimes with highly unexpected and 
extreme characteristics. More than 2900 of the identified or associated high energy (HE, $>100$ MeV) sources in 
the Fermi-LAT 8-year Point Source List (FL8Y)\footnote{http://fermi.gsfc.nasa.gov/ssc/data/access/lat/fl8y/},
are active galactic nuclei (AGN) of the blazar class. In the very high energy (VHE, $>100$ GeV) domain the detection 
of about 70 AGN is currently summarised in the TeVcat catalog\footnote{http://tevcat.uchicago.edu}. 
Again, most of these sources are of the blazar type, i.e. radio-loud AGN such as BL Lac objects where the jet is thought 
to be inclined at very small viewing angles $i$ to the line of sight. This results in substantial Doppler-boosting of their 
intrinsic jet emission, $S(\nu)=D^{a}S'(\nu')$ where $D=1/[\Gamma_b(1-\beta_b\cos i)]$ denotes the Doppler factor 
and $\Gamma_b=(1-v_b^2/c^2)^{-1/2}$ the jet bulk Lorentz factor and typically $a\geq 2$, privileging their detection on 
the sky. Nevertheless, non-blazar AGN such as Radio Galaxies (RGs), while less occurrent, have in the meantime solidly 
emerged as a new gamma-ray emitting source class as well. With their jets misaligned and associated Doppler boosting 
effects modest, they enable unique insights into often hidden regions and processes. This review aims at a compact 
summary of their properties and highlights their role in facilitating theoretical progress in AGN physics.\\ 
The unification model of radio-loud AGNs postulates that RGs are viewed at a substantial inclination $i$ to the jet 
axis so that the broad-line optical emitting regions become obscured by a dusty component ("torus" or warped 
disk) in Narrow Line RGs (NLRGs) such as in Cen~A or M87 \cite{1989ApJ...336..606B,1993ARA&A..31..473A}. 
Depending on their radio structure, RGs have early on been divided into Fanaroff-Riley I and II sources (FR I, FR II) 
\cite{1974MNRAS.167P..31F}, the former one (FR I) encompassing lower radio luminosity, edge-darkened sources 
and the latter one (FR II) higher luminous, edge-brightened sources where the radio lobes are dominated by bright
hot spots. Various considerations suggest that the high-power FR II sources might be accreting in a "standard" 
(geometrically thin, optically thick) mode, while most FR I sources are probably supported by a radiatively inefficient 
accretion flow (RIAF) \citep{Ghisellini2001,Wang2003}.\\
The general relationship between the blazar and RG class is complex. Urry \& Padovani~(1995) have described 
BL Lacs as beamed FR I RGs \cite{1995PASP..107..803U}, though evidence exists that the parent population of BL 
Lac objects contains both FR I and FR II sources \cite{1992AJ....104.1687K,2012A&AT...27..557A}. A more detailed 
view might be to posit that X-ray loud BL Lacs (mostly HBLs, peaking in UV/X-rays) are preferentially associated 
with FR I, while radio-loud BL Lacs (mostly LBL, peaking in the infrared) could show a mixture of FR I and FR II 
morphologies \cite{2000AJ....120.1626R}.

\section{Radio Galaxies as VHE emitters - Experimental Status}
\begin{figure}[h]
  \centering
  \includegraphics[width=350pt]{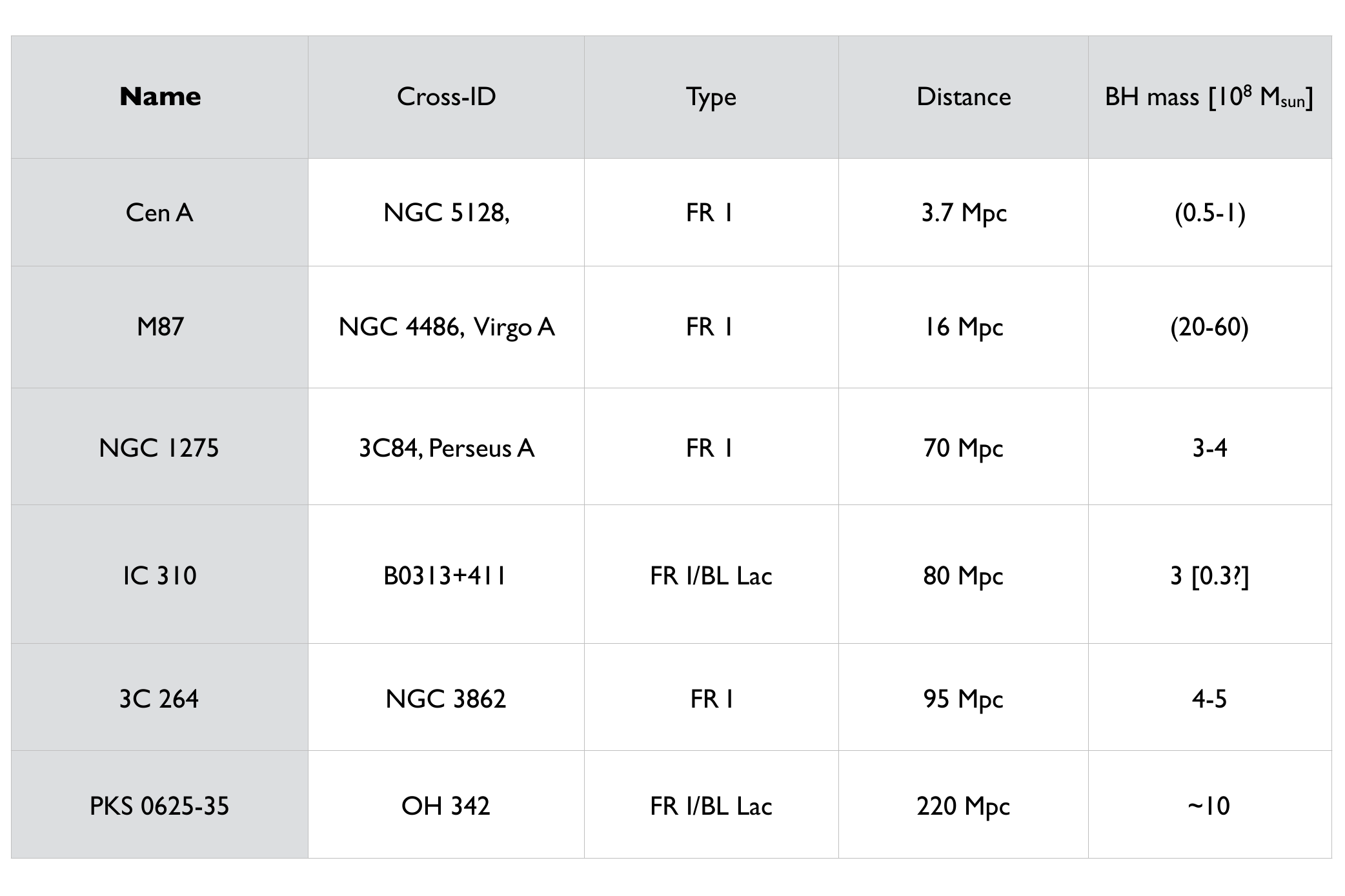}
  \caption{Radio galaxies reported at VHE energies, including estimates for their black hole masses. Cross-IDs give 
  their alternative source identifications. Two sources, IC~310 and PKS~0625-35, may be of a transitional type.}\label{rg_list}
\end{figure}

In the HE range Fermi-LAT has detected about 20 RGs  \citep[e.g.,][]{Narek2018}. Six of them are also known
as VHE emitters (see Fig.~\ref{rg_list}), including M87 ($d\sim 16$ Mpc), the first extragalactic source detected
at VHE energies, and Cen A, the nearest ($d\sim 4$ Mpc) AGN to us.
\noindent

This emergence of RGs as a new VHE emitting source class is particularly interesting. Given the substantial 
misalignment of their jets ($i>10^{\circ}$), RGs are commonly thought to be characterized by rather modest Doppler 
boosting only (bulk Doppler factor $D\leq$ a few). If, following simple unification considerations, the nuclear 
emission of FR~I type RGs is interpreted as "misaligned BL Lac type" (i.e., of a jet-related, homogeneous 
synchrotron self-Compton [SSC] origin, yet with small Doppler factor)\citep{Chiaberge2001}, only a few sources 
should become detectable at GeV energies (which seems indeed to be the case), and almost none at TeV 
energies. The discovery of RGs as a new VHE emitting source class thus points to a more complex situation,
and promises new insights into some of the fundamental (and often hidden) non-thermal processes in 
$\gamma$-ray emitting AGN. The following aims to provide a short summary of the experimental source 
characteristics:

\subsection{PKS 0625-354}
The detection of VHE emission from {\it PKS 0625-354} ($z=0.055$) above 200 GeV (at a level of $\sim6\sigma$ 
in 5.5 h of data) has been recently reported by H.E.S.S. \citep{HESS2018_PKS0625}. No significant variability 
is found in the data. The VHE spectrum extends up to $\sim 2$ TeV and is compatible with a rather steep power 
law of photon index  $\Gamma_{\gamma} \sim -2.8\pm0.5$. The VHE power is moderate with an apparent isotropic 
luminosity of the order of $L_{\rm VHE} \sim 5 \times 10^{42}$ erg/s. Both leptonic and hadronic SED interpretations
seem possible \citep{HESS2018_PKS0625}.\\
PKS 0625-354 is thought to harbour a black hole of mass $M_{\rm BH} \sim 10^9 M_{\odot}$ \citep{Mingo2014} 
that is probably accreting in an inefficient mode. The source is known as a low excitation line radio-loud AGN, but 
being a transitional FR I / BL Lac object its proper classification has been debated. Recent findings are favouring 
its classification as a BL Lac object with non-modest Doppler boosting \citep{Wills2004,Ramos2011,Mueller2013}. 
It inclusion in the list of (misaligned) "radio galaxies" may thus have to be re-considered, limiting possible 
inferences as to the physical origin of its non-thermal emission based on current data 
\citep[e.g.,][]{Fuk2015,HESS2018_PKS0625}.

\subsection{3C~264}
The most recent addition to the RG list has been the FR I source {\it 3C~264} ($d\sim 95$ Mpc) seen by VERITAS 
(with a significance level of $\sim5.4\sigma$ in 12 h of data) \citep{Muk2018}. Given an estimated black hole 
mass $M_{\rm BH} \sim 5\times10^8 M_{\odot}$ \cite{Ruiter2015}, the VHE luminosity appears to be moderate 
($\sim 1\%$ of the Crab Nebula) with an isotropic equivalent $L(>300$ GeV$) \sim 10^{42}$ erg/sec. 3C~264 
has been included in the 3FHL catalog, that lists Fermi-LAT sources which are significantly detected above 10 GeV 
\cite{Ajello2017}. The reported VHE flux level seems roughly compatible with a simple power law extrapolation 
based on the 3FHL results (FHL photon index of $-1.65\pm0.33$). There are indications, though, that the VHE 
spectrum is relatively hard (when compared to other VHE sources) with a photon index close to $\Gamma_{\gamma}
\sim -2.3$.  The source shows a low, weakly variable VHE flux along with some month-scale variations. While 
3C 264 is known for rapidly evolving knot structures in its jet up to some hundred of parsecs \citep{Meyer2015}, 
no major knot activity has been observed around the time of the VERITAS observations. Given the previously 
noted unclear classification of PKS 0625-354, 3C~264 may be the most distant RG detected at VHE so far.

\subsection{NGC 1275} 
{\it NGC 1275} (3C~84), the central  Perseus cluster RG at a distance of $\sim 70$ Mpc, has been detected at 
VHE energies above 100 GeV by MAGIC, initially (based on data between 2009-2011) at moderate flux levels 
($\sim 3\%$ of the Crab Nebula) and with a very steep VHE spectrum (photon index of $\Gamma_{\gamma}
\sim-4.1$ if characterized by a power law) extending up to $\sim650$ GeV \citep{Aleksic2012, Aleksic2014}. 
When HE (Fermi-LAT) and VHE data are combined, the average ("quiescent") $\gamma$-ray spectrum appears 
compatible with either a log-parabola or a power-law with a sub-exponential cut-off, suggestive of a common 
physical origin and of a peak or cut-off around several GeV. More recently, MAGIC has reported the detection 
of strong VHE activity with flux levels increased by up to a factor of 50 around December 31, 2016 and 
January 1, 2017 (reaching $\sim1.5$ of the Crab Nebula or an isotropic equivalent $L_{VHE} \sim 10^{45}$ 
erg/sec)\citep{MAGIC2018}. 
\begin{figure}[th]
  \centering
  \includegraphics[width=300pt]{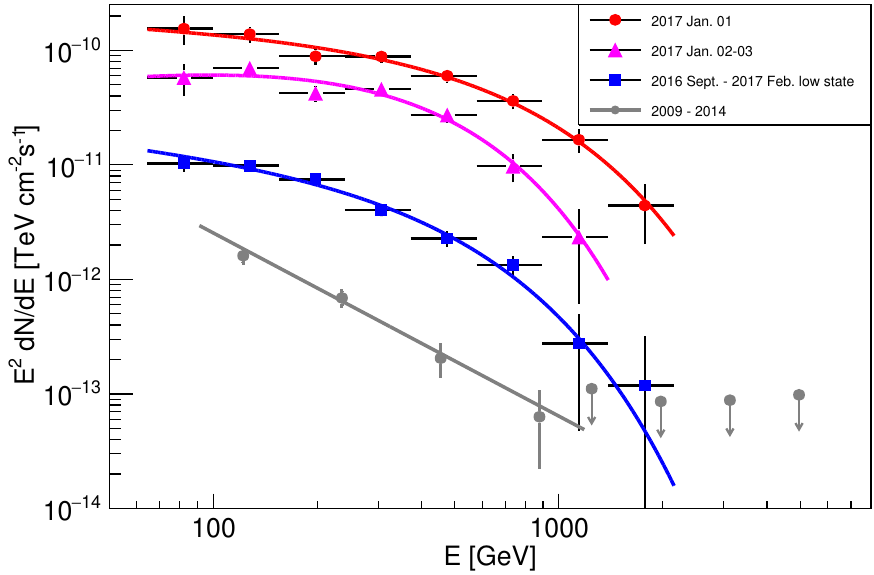}
  \caption{The VHE SEDs of NGC 1275 as measured by MAGIC during different periods. Significant curvature is 
  evident, suggestive of an exponential cut-off around a few hundred GeV. For comparison the averaged spectrum 
  based on observations in 2009 to 2014 is shown in grey. From Ref.~\citep{MAGIC2018}.}
  \label{NGC1275_spectrum}
\end{figure}
Significant day-scale variability has been observed, with the flux doubling timescales as short as $\Delta t_{\rm obs}
\sim10$ h. The VHE SED measured up to $> 1$ TeV shows a curved shape (cf. Fig.~\ref{NGC1275_spectrum}), 
compatible with an exponential cut-off around a few hundred GeV. The possibility of a joined HE-VHE fit along 
with day-scale variability, suggests that the HE-VHE emission originates in a (possibly, single) compact zone.
The physical nature of this emission is not yet clear, though magnetospheric processes have been favoured 
over mini-jets- and jet-cloud-interaction scenarios \cite{MAGIC2018}.\\ 
The central engine in NGC 1275 hosts a black hole of mass $M_{\rm BH} \sim(3-4) \times 10^8 M_{\odot}$ 
\citep{Wilman2005,Wu2011} and exhibits a pc-scale radio jet orientated at $i\sim 30-60^{\circ}$ \citep{Walker1994,
Fujita2017}. Its inferred jet power is of the order of $L_j\sim (0.5-1) \times 10^{44}$ erg/sec \citep{Wu2011,
Fujita2016}. The high ratio $L_{VHE}/L_j \sim 10$ thus raises questions for a magnetospheric origin of the 
gamma-ray flare emission, cf. \citep{Greg2018} (see also below), unless strong short-term magnetic flux
increase would occur. On the other hand, a homogeneous SSC interpretation, assuming the sub-pc scale jet to 
be weakly misaligned ($i\simlt20^\circ$), would be in tension with the inferred jet inclination on pc-scales. This 
could perhaps be alleviated if the emitting component would e.g. follow a non-straight trajectory that relaxes with 
distances, and/or if the jet has some internal structure (e.g., spine-shear) allowing for multiple contributions and 
a more complex inverse Compton interplay \citep{Tavecchio2014}. Opacity constraints may pose a severe problem, 
though (see below). At the moment detailed modelling seems required before firm conclusions can be drawn.

\subsection{Centaurus~A}
\begin{figure}[h]
  \centering
  \includegraphics[width=320pt]{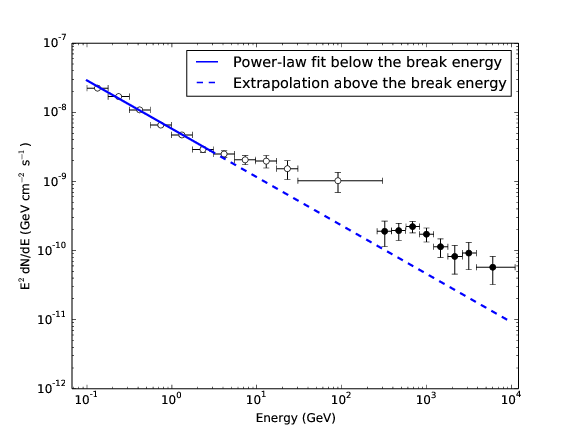}
  \caption{The gamma-ray core spectrum of Cen A above 100 MeV based on 8 yr of Fermi-LAT and more 
  than 200h of H.E.S.S. data. The spectrum shows an unusual spectral hardening at $E_b\simeq2.8$ GeV, 
  with photon index changing by $\sim 0.4\pm0.1$ (assuming a broken power law), see Ref.~\citep{HESS2018_CenA} 
  for details. This spectral feature is most naturally attributed to a second emission component that emerges 
  towards highest energies and that allows to smoothly connect the HE emission (above $E_b$) with the 
  VHE one.}
  \label{CenA_spectrum}
\end{figure}

As the nearest AGN ($d\simeq 3.7$ Mpc) Centaurus A ({\it Cen~A}) belongs to the best studied extragalactic sources. 
Its central engine hosts a black hole of mass $(0.5-1) \times 10^8 M_{\odot}$ \citep[e.g.,][]{Neumayer2010} and 
emits (assuming a quasar-type SED) a bolometric luminosity of $L_{\rm bol} \sim 10^{43}$ erg/sec \citep{Whysong2004}. 
This is much less than the expected Eddington luminosity $L_{\rm Edd}$ and suggests that accretion in its inner 
part might occur in a radiatively inefficient mode \citep{Meisenheimer2007,Fuerst2016}. 
At radio frequencies Cen~A has revealed a peculiar morphology including a compact radio core, a sub-pc scale 
jet and counter-jet, a one-sided kpc-scale jet and inner lobes, up to giant outer lobes with a length of hundreds of 
kiloparsec. VLBI observations indicate that Cen~A is a "non-blazar" source with its inner jet misaligned by $i\sim 
(12-45)^{\circ}$ based on TANAMI jet-counter jet flux ratio measurements, and characterized by moderate bulk 
flow speeds in the radio band of $u_j  < 0.5$ c only \citep[e.g.,][]{Mueller2014}.\\  
At VHE energies Cen~A has been the second RG detected by H.E.S.S.  \citep{Aharonian2009}. A recent, updated 
analysis based on more than 200h of data shows that the VHE emission extends from 250 GeV up to $\sim6$ TeV 
and is compatible with a single, rather hard power-law of photon index $\Gamma_{\gamma} \simeq -2.5\pm0.1$ 
\citep{HESS2018_CenA}. The source is relatively weak with an equivalent apparent isotropic luminosity of 
$L(>250$ GeV$) \simeq (1-2) \times 10^{39}$ erg/s. No significant VHE variability has been found, neither on 
monthly or yearly timescales, so that an extended origin or contribution (i.e., within the angular resolution $\sim 
0.1^{\circ}$ of H.E.S.S., corresponding to $\sim 5$ kpc) of the VHE emission cannot per se be discarded.\\
At HE energies, both the core region (i.e., within $\sim 0.1^{\circ}$) and the giant lobes of Cen~A have been 
detected by Fermi-LAT \citep{Abdo2010, Abdo2010_Sci, Yang2012, Sun2016}. Results concerning the later
are indicating that HE lobe emission substantially extends beyond the radio maps. The HE emission of the 
lobes (which is most likely due to leptonic IC-CMB and IC-EBL, possibly with some additional hadronic pp) 
is of a particular interest as it provides model-independent information about the spatial distribution of the 
non-thermal electrons. Fermi-LAT has by now reported extended HE emission from only two RGs, Cen A 
and Fornax A ($d\sim 20$ Mpc) \citep{Ackermann2016}.
The core region of Cen~A, on the other hand, was initially detected up to 10 GeV (at a level of $4\sigma$) 
based on ten months of data, with the HE spectrum at that time seemingly compatible with a single power law 
with photon index $\Gamma_p = -2.67 \pm 0.1$. While this HE power index is very close to the VHE one, a 
simple extrapolation of the HE power-law was soon found to under-predict the fluxes measured at TeV energies. 
The comparison was based on non-simultaneous HE and VHE data, but the absence of variability in both 
energy bands suggested that the discrepancy might be real. Refined analyses based on larger data sets 
have in the meantime found intriguing evidence for an unusual spectral hardening of the core spectrum by 
$\Delta \Gamma \sim 0.5$ around a few GeV \citep{Sahakyan2013,Brown2017}. The most recent analysis, 
involving contemporary VHE and HE data, finds (at a level of $4\sigma$) that the HE spectral index changes 
around $E_b\simeq 2.8$ GeV from $\Gamma_{\gamma} \simeq-2.7$ (below $E_b$) to about $\simeq -2.3$ 
(above $E_b$), respectively \citep{HESS2018_CenA}, see Fig.~\ref{CenA_spectrum}.\\
\noindent
For AGN spectral steepening at gamma-ray energies is a familiar feature that can be related to classical constraints on 
the acceleration and radiation efficiencies. The observed spectral hardening in Cen~A is unusual in this regard; in a 
"misaligned BL Lac approach" it is best understood as related to the presence of an additional emission component 
beyond the conventional single-zone SSC-contribution that often satisfactorily describes the SED in blazars. Apart from 
circumstantial evidence for the blazar Mkn~501 \citep{Neronov2012,Shukla2016}, Cen~A is the first source where spectral 
results provide clear evidence for the appearance of a physically distinct component above a few GeV. Unfortunately, 
Cen~A is a rather weak $\gamma$-ray emitting source, which significantly limits the possibilities to further probe its
variability characteristics, particularly above the break. This makes it difficult to observationally disentangle the 
true nature of the second component with current data.\\
In principle a variety of different (not mutually exclusive) interpretations as to its astrophysical origin are conceivable. 
Related proposals in the literature operate on different scales (from a few $r_g$ to several kpc) and 
include: (i) (rotational) magnetospheric models that are based on leptonic inverse Compton (IC) processes in an 
under-luminous accretion environment \citep{Rieger2009,Rieger2011}, (ii) inner (parsec-scale and below) jet 
scenarios that invoke differential IC scattering in a stratified jet  \citep{Ghisellini2005}, multiple SSC-emitting 
components moving at different angles to the line of sight \citep{Lenain2008} or photo-meson (p$\gamma$) 
interactions of UHE protons in a strong photon field \citep{Kachelriess2010,Sahu2012,Petropoulou2014} along with 
lepto-hadronic combinations \citep{Reynoso2011, Cerruti2017}; alternatively, the hardening could be related to 
$\gamma$-ray induced pair-cascades in a strong disk photon field \citep{Sitarek2010}, a dusty torus-like region
 \citep{Roustazadeh2011} or the overall host photon field \citep{Stawarz2006}. Moreover, the limited angular resolution 
 of Fermi-LAT and H.E.S.S. ($\sim 5$ kpc), and the fact that no significant statistical evidence for variability has been 
 found so far, also allows for (iii) scenarios where the emission arises on larger scales; extended scenarios in this 
 context include the interaction of energetic protons with ambient matter (pp) in its kpc-scale region \citep{Sahakyan2013}, 
the overall $\gamma$-ray contribution of a supposed population of millisecond pulsars \citep{Brown2017}, or the IC 
contribution by its kpc-scale jet via up-scattering off various photon fields (e.g., host galaxy starlight or CMB) 
\citep{Stawarz2003, Hardcastle2011}, up to more extraordinary explanations invoking the self-annihilation of dark 
matter particles of mass $\sim 3$ TeV within a central dark matter spike \citep{Brown2017}.\\
While, given current knowledge, not all of these models are equally likely, and all of them come with some challenges 
\cite[see e.g.,][]{Rieger2017}, further observational input (such as evidence for VHE variability or extension, the latter 
now possibly been seen \citep{Sanchez2018}) is needed to better constrain them and help disclosing the real nature 
of this new component.

\subsection{IC 310}
\begin{figure}[h]
  \centering
  \includegraphics[width=300pt]{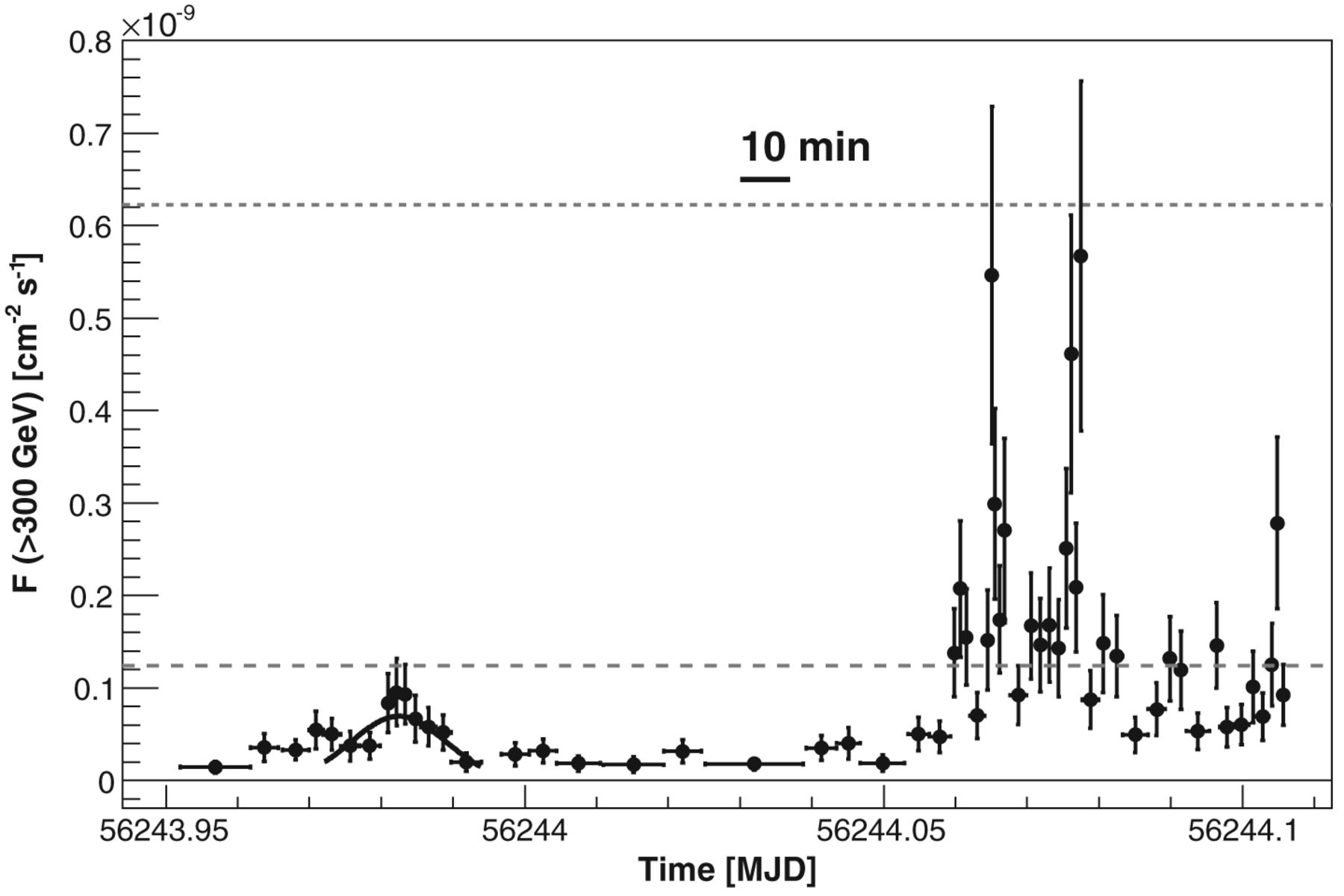}
  \caption{VHE light curve of IC 310 above 300 GeV as observed by MAGIC during November 12/13, 2012. Rapid VHE 
  variability on doubling timescale well below 10 min is apparent in the light curve. The two gray lines indicate  flux levels 
  of 1 and 5 Crab units, 
 respectively. From Ref.~\citep{Aleksic2014a}.}\label{IC310}
\end{figure}

The Perseus Cluster RG {\it IC~310}, located at a distance of $d\sim 80$ Mpc (z=0.019), has received particular attention 
in recent times. The source, originally detected by MAGIC during a campaign in 2009-2010 \citep{Aleksic2010}, has shown 
extreme VHE variability during a strong flare in November 2012, revealing VHE flux variations on timescales as short as 
$\Delta t\simeq 5$ min \citep{Aleksic2014a}, see Fig.~\ref{IC310}. The 2012 VHE flare spectrum appears compatible with a 
single, hard power law of photon index $\Gamma_{\gamma} \simgt -2$ (and possibly as low as $\sim-1.5$) over a range from 70 
GeV to 8.3 TeV, with no indications of any internal absorption \citep[see also][]{Ahnen2017}. The source can reach high VHE 
flux levels, corresponding to an isotropic-equivalent luminosity of $L_{VHE} \simeq 2 \times 10^{44}$ erg\,s$^{-1}$. IC~310 
is commonly believed \citep[e.g.,][]{Aleksic2014a} to harbour a black hole of mass $M_{BH} \simeq 3 \times 10^8 M_{\odot}$ 
\citep[but see also Ref.][for a ten times smaller estimate]{Berton15} and has for some time been classified as a head-tail RG. 
The apparent lack of jet bending along with more recent indications for a one-sided pc-scale radio jet inclined at $i\simlt 
38^{\circ}$ suggests, however, that IC~310 is a transitional source at the borderline dividing low-luminosity RGs and BL Lac 
objects \citep{Kadler2012}.
The extreme VHE variability along with the high VHE power ($\simgt L_{\rm Edd}/200$) and the hard $\gamma$-ray spectrum 
are surprising findings for a misaligned source. Based on a variety of considerations, including the orientation of its jet 
(probably $i\sim [10-20]^{\circ}$) as well as kinetic jet power and timing constraints, \citet{Aleksic2014a} have disfavoured 
several alternative models for rapid VHE variability such as magnetic reconnection \citep[e.g.,][]{Giannios2013} or jet-cloud/star 
interaction \citep[e.g.,][]{Barkov2012}. This inference is, however, less robust as has been shown later on \citep[cf.][for 
details]{Aharonian2017}. 
Nevertheless, the fact that the VHE flux varies on timescales $\Delta t$ much shorter than the light travel time across the 
black hole horizon, $r_g(3\times10^8 M_{\odot})/c = 25$ min, has been interpreted as evidence for the occurrence of gap-type 
particle acceleration on sub-horizon scales, i.e. in unscreened electric field regions ("gaps") of height $h\simeq 0.2 r_g$ 
\citep[e.g.,][]{Aleksic2014a,Hirotani2016}. Questions concerning such an interpretation are related to the fact that the characteristic 
VHE power of a (steady) gap scales with the jet power, $L_{VHE} \sim L_{\rm j} (h/r_g)^a$, $a=2-4$ \citep{Greg2018}, the latter 
of which is known to be rather modest on average for IC~310, i.e. $L_j \sim 10^{43}$ erg\,s$^{-1}$ \citep[cf. also,][]{Sijbring1998}. 
Unless strong (short-term) magnetic flux increases would occur, the expected gap output would under-predict the VHE fluxes 
measured during the flaring state. IC~310 has subsequently (after November 2012) shown a rather low TeV emission state with a 
steeper spectrum ($\Gamma \sim -2.4$) measured up to $\sim 3$ TeV and with little evidence for variability. The multi-wavelength
SED during this state appears to be satisfactorily reproducible with a one-zone SSC model using parameters that are comparable 
to those found for other misaligned, $\gamma$-ray emitting AGN \citep{Ahnen2017}.

\subsection{M87}
The Virgo Cluster galaxy {\it M87} (NGC 4486) has been the first extragalactic source detected at VHE energies 
\cite{Aharonian2003}. Classified as low-excitation, weak-power FR I source, M87 hosts one of the most massive 
black holes of $M_{\rm BH} \simeq (2-6)\times 10^9\,M_{\odot}$ \citep[e.g.,][]{Walsh2013}, and is thought to be 
accreting in a radiatively inefficient (RIAF) mode \cite{Reynolds1996}. Given its proximity at a distance of $d\simeq 
16.4$ Mpc \citep{Bird2010} and its large mass-scale $r_g$, M87 has become a prominent target to probe jet formation 
scenarios with high-resolution radio observations down to scales of tens of gravitational radii \citep[e.g.,][]{Doeleman2012,
Kino2015, Akiyama2015,Hada2016,Mertens2016}. Its sub-parsec scale radio jet appears misaligned by an angle $i \sim 
(15-25)^{\circ}$ and shows a rather complex structure, seemingly compatible with a slower, mildly relativistic ($\beta 
\sim 0.5$c) layer and a faster moving, relativistic spine ($\Gamma_b \sim 2.5$)\citep[see e.g.,][]{Mertens2016}. 
Indications of a parabolic jet shape suggest that the jet initially experiences some external confinement as by a disk 
wind \citep{Globus2016}. In general the inferred jet seeds and inclinations are consistent with rather modest Doppler 
factors $D\simlt$ a few (for review, see e.g. \citep{Rieger2012}). 

\begin{figure}[h]
  \vspace*{-0.4cm}
  \centering
  \includegraphics[width=350pt]{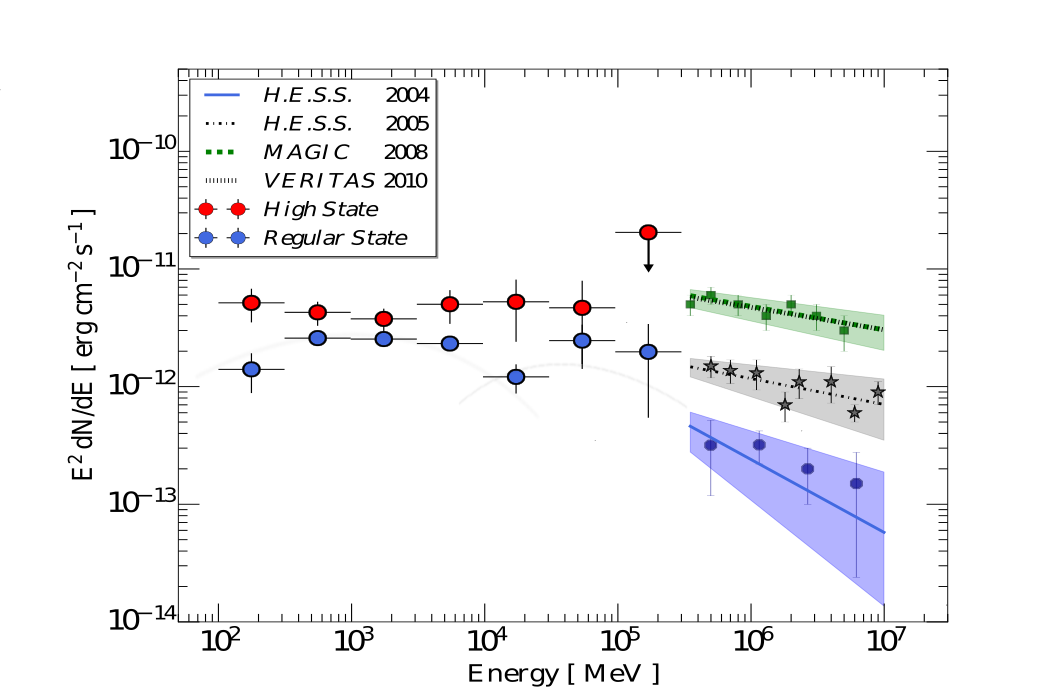}
  \caption{Gamma-ray SED for M87 based on $\sim8$ yr of Fermi-LAT data including the different observed VHE
  states. The average ("regular") spectrum shows a break in the SED around $\sim 10$ GeV, suggestive of an 
  additional HE component. The break appears masked in the "high state" by flaring above $\sim10$ GeV. The 
  current situation now in principle allows for a smooth connection of HE and VHE states. This suggests that the
  nature of this additional component is constrained by the observed VHE variability. The light grey curves 
  for the two components are intended to guide the eyes only. Following Ref.~\citep{Faical2018}.}\label{M87}
\end{figure}

At VHE energies, M87 is well known for its rapid day-scale variability (flux doubling time scales $\Delta t_{\rm obs} 
\sim 1$ d) during active source states, and a rather hard, featureless photon spectrum compatible with a single 
power law (of index $\Gamma_{\gamma} = -2.2 \pm 0.2 $ in high, and somewhat steeper $\Gamma_{\gamma}
\sim -2.6$ in low states) extending from $\sim 300$ GeV up to $\sim 10$ TeV \citep{Aharonian2006,Albert2008,
Acciari2009,Aliu2012,Abramowski2012}. Both, the observed rapid VHE variability and the hard VHE spectrum are 
remarkable features for a misaligned AGN, and reminiscent of those seen in IC~310. Based on the first 10 months 
of data, Fermi-LAT has reported HE gamma-ray emission from M87 up to 30 GeV \citep{Abdo2009} with a photon 
spectrum then seemingly compatible with a single power-law of index $\Gamma_{\gamma}=-2.26 \pm 0.13$ and 
comparable to the one(s) in the VHE high states. Nevertheless, a simple extrapolation of this HE power-law to the 
VHE regime turned out to be insufficient to account for the flux levels measured during the TeV high states (up to 
equivalent levels of $L(>350~\mathrm{GeV}) \sim 5\times 10^{41}$ erg/s, e.g. \citep{Aliu2012}), suggesting that the 
high states might be accompanied by the emergence of an additional component \citep{Rieger2012}.
No evidence for significant flux variations (down to timescales of 10 days) has been found during these early HE 
observations, though on experimental grounds the occurrence of shorter-timescale variations cannot per se be 
excluded. Similar spectral results have been reported in the 3FGL catalog (4 yr of data), with the HE spectrum below 
10 GeV compatible with a single power-law of $\Gamma_{\gamma}=-2.04\pm0.07$ \citep{Acero2015}, but with 
indications for a possible change above 10 GeV. The most recent analysis based on $\sim 8$ yr of Fermi-LAT data 
reports evidence for month-type HE variability and indications for excess emission over the standard power-law 
model above $\sim10$ GeV, similar to earlier findings in Cen~A \citep{Faical2018}, see also Fig.~\ref{M87}. When 
viewed in an HE-VHE context, these findings are most naturally explained by an additional emission component 
that dominates the highest-energy part of the spectrum and allows for a smooth HE-VHE spectral connection. As 
the HE spectrum extends to about 100 GeV without indications for a cut-off and the VHE thresholds reach down to 
about 200 GeV, variability seen with high statistics at VHE can be used to constrain the nature of this additional 
component. This contrasts with Cen~A where no significant VHE variability has been found yet. For M87 current 
findings do support proposals in which the emission arises on innermost jet scales and below.

\begin{figure}[h]
  \centering
   \includegraphics[width=380pt]{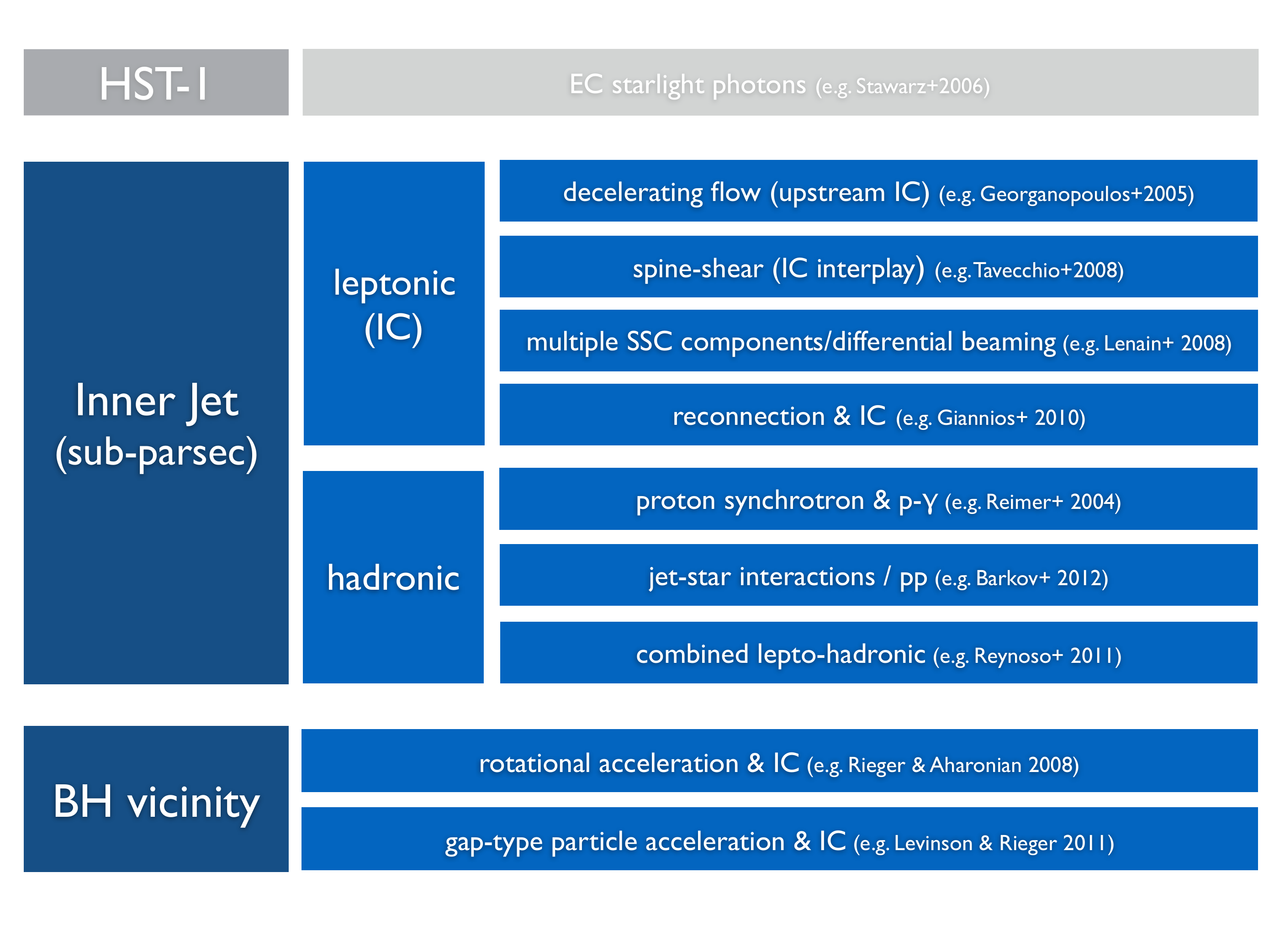}
  \caption{Possible scenarios for the origin of the variable VHE emission in M87 with exemplary references. 
  Day-scale variability favours models on scales of the inner jet and below. "IC" refers to inverse Compton 
  radiation.}\label{interpretations}
\end{figure}

Light travel time arguments in fact point to a compact VHE emission region ($R < c\Delta t_{\rm obs} D$) in M87
of a size comparable to the Schwarzschild radius $r_s=(0.6-1.8) \times10^{15}$ cm of its black hole. Similar as 
for Cen~A, a variety of models have been introduced to account for this, cf. Fig.~\ref{interpretations} for an
exemplary illustration (see Refs.~\citep{Tavecchio2008,Lenain2008,Giannios2010,Reimer2004,Barkov2012,
Georganopoulos2005,Reynoso2011,Rieger2008,Levinson2011}). The interested reader is referred to 
Refs.~\citep{Rieger2012,Rieger2012b} for a more detailed description and discussion of them.\\
M87 has been repeatedly active over the past ten years, with VHE high states being detected in 2005, 2008 
and 2010, and an elevated one (flux levels 2-3 times higher than average) in 2012. Interestingly, during all high 
states, day-scale VHE variability has been found. The 2012 monitoring data by VERITAS do not reveal a
bright flare, but the light curve indicates VHE variability on timescales of (at least) weeks, suggesting that the 
often called "quiescent" state also shows some longterm evolution \citep{Beilicke2012}, cf. also \citep{Faical2018}. 
No major VHE flare has been seen since then, though hints for day-scale variability in 2013 have been reported 
\citep{Bangale2015}. As the angular resolution of current VHE instruments is limited (to scales of $\sim25$ 
kpc for M87), coordinated VLBI radio observations, capable of probing down to scales of tens of gravitational 
radii, have been performed during the 2008, 2010 and 2012 high VHE states. These results indicate that the
TeV emission is accompanied by (delayed) radio core flux enhancements, supporting proposals that the VHE 
emission originate at the jet base very near to the black hole \cite{Acciari2009,Hada2012,Hada2014,Akiyama2015}. 
The radio--VHE correlation along with the required compactness of the VHE zone have served as a strong 
motivation to explore plasma injection via gap-type magnetospheric processes close to the black hole 
\citep{Levinson2011,Broderick2015,Ptitsyna2016}. 
Concerning its rapid VHE variability M87 shows some similarities with IC\,310, though its associated VHE 
luminosity output is by more than a factor of 100 smaller than the one for IC\,310.

\section{Models for the HE-VHE $\gamma$-ray emission}
In the following we shortly describe and comment on some recent theoretical trends and developments 
concerning the origin of the VHE emission beyond conventional jet (one-zone) synchrotron self-Compton 
(SSC) models. We distinguish between scenarios aimed at addressing the highly variable VHE part (as 
in M87) and those focusing on an apparently steady VHE part (as for Cen~A).

\subsection{Variable VHE and black hole gap models}
The activation of  Blandford-Znajek (BZ) outflows requires continuous injection of plasma in the magnetospheric 
region enclosed between the inner and outer light cylinders, the origin of which is yet an open issue.  
To fully screen out the magnetosphere, the plasma injection rate must be sufficiently high to maintain the density 
everywhere in the magnetosphere above the Goldreich-Julian (GJ) value.  
Whether the dense accretion flow surrounding the black hole can provide sufficient charges for complete screening 
is unclear at present; direct feeding seems unlikely, as charged particles would have to cross magnetic field lines 
on a timescale much shorter than the accretion time in order to reach the polar outflow. Plasma injection by virtue 
of macroscopic instabilities, that might lead to re-arrangement of the magnetic field configuration, is a possibility, 
however, the growth time of such instabilities may be much longer than the gap evacuation time, $\sim r_g/c$, 
and it is quite likely that even if occasional injection of plasma into  the casual section of the magnetospheric 
does occur, the plasma density may not be sustained above the required level at all times.
An alternative charge supply mechanism is pair creation on magnetic field lines via annihilation of MeV photons 
that emanate from a RIAF during low accretion states, or from a putative accretion disk corona during intermediate 
states.     
To estimate the $\gamma$-ray luminosity required for complete screening we note that the density of pairs thereby 
created is roughly $n_\pm\simeq \sigma_{\gamma\gamma}n^2_\gamma r_g/3$ \citep{Levinson2011}, where 
$\sigma_{\gamma\gamma}$ is the pair production cross section, $n_\gamma\simeq 10^{22}\, 
m^{-1}\tilde{R}_\gamma^{-2} l_\gamma$  cm$^{-3}$ is the density of MeV photons,  $l_\gamma=L_\gamma/L_{Edd}$ 
and $R=\tilde{R} r_g$ are, respectively, the Eddington ratio and radius of the radiation source, and $m=M/M_\odot$ 
is the black hole mass in solar mass units.  This should be compared with the GJ density, $n_{GJ}=\Omega B/(2\pi e c)
=2\times10^{11} B_8(\Omega/\omega_H)m^{-1}$ cm$^{-3}$, expressed here in terms of  the angular velocity of 
magnetic surfaces $\Omega$,  the angular velocity of the black hole $\omega_H\simeq c/2r_g$,  the strength of 
the magnetic field near the horizon $B=10^8 B_8$ Gauss, and  the magnitude of the electron charge $e>0$.  
The requirement $n_\pm > n_{GJ}$ implies
\begin{equation}
l_\gamma > 10^{-3} B_8^{1/2}(\Omega/\omega_H)^{1/2} (\tilde{R}/30)^2.
\end{equation}
For sources accreting in the RIAF regime, the gamma-ray luminosity can be related to the accretion rate, albeit with a 
large uncertainty, using an ADAF model. The strength of the magnetic field near the horizon also scales with the 
accretion rate, roughly as $B\simeq 10^9 (\dot{m}/m)^{1/2}$ G, where $\dot{m}=\dot{M}/\dot{M}_{Edd}$, and 
$\dot{M}_{Edd}=10L_{Edd}/c^2$. These two relations can be combined to yield the critical accretion rate below 
which the magnetosphere is expected to be starved,
\begin{equation}
\dot{m} <4\times10^{-3} m^{-1/7}.
\label{eq:dotM_crit}
\end{equation}
For M87, where $m=6\times10^9$ (up to a factor of two), this implies starvation at $\dot{m}\simlt 10^{-4}$, which 
seems to be above its inferred accretion rate.
In sources accreting well above the critical ADAF rate, the accretion flow is anticipated to be cold, and the emission 
spectrum is unlikely to extend to energies above the electron mass. Gamma-rays may, nonetheless, originate from 
a tenuous corona, if present. No reliable constraints on the spectrum and luminosity of this coronal component have 
been imposed thus far. In principle, it could be that in sources that accrete at relatively high rates the magnetic field 
is much higher than in RIAF sources, while the gamma-ray luminosity (of the corona) is smaller. If indeed true, it could 
mean that gap emission in such objects may be more intense than in RIAF sources.
Equation (\ref{eq:dotM_crit}) has been employed to show \citep{Levinson2011,Hirotani2016,Greg2018} that under 
conditions likely to prevail in many stellar and supermassive black hole systems, the annihilation rate of disk photons is 
insufficient to maintain the charge density in the magnetosphere at the GJ value, giving rise to formation of spark 
gaps \cite{Blandford1977,Beskin1992, Hirotani1998,Levinson2000}. It has been further pointed out \citep{Neronov2007, 
Rieger2011, Levinson2011, Hirotani2016,Hirotani2016b,Lin2017,Greg2018} that the gap activity may be imprinted 
in the high-energy emission observed in these sources, whereby the variable TeV emission detected in M87 
\citep{Aharonian2006, Acciari2009} and IC310 \citep{Aleksic2014a} was speculated to constitute examples of the 
signature  of magnetospheric plasma production on horizon scales \citep{Levinson2000,Neronov2007,Levinson2011,
Ptitsyna2016,Hirotani2016, Greg2018}. In what follows we provide a concise overview of recent black hole gap models.

\subsubsection{Stationary gap models}
\begin{figure}[h]
\centering
\includegraphics[width=7cm]{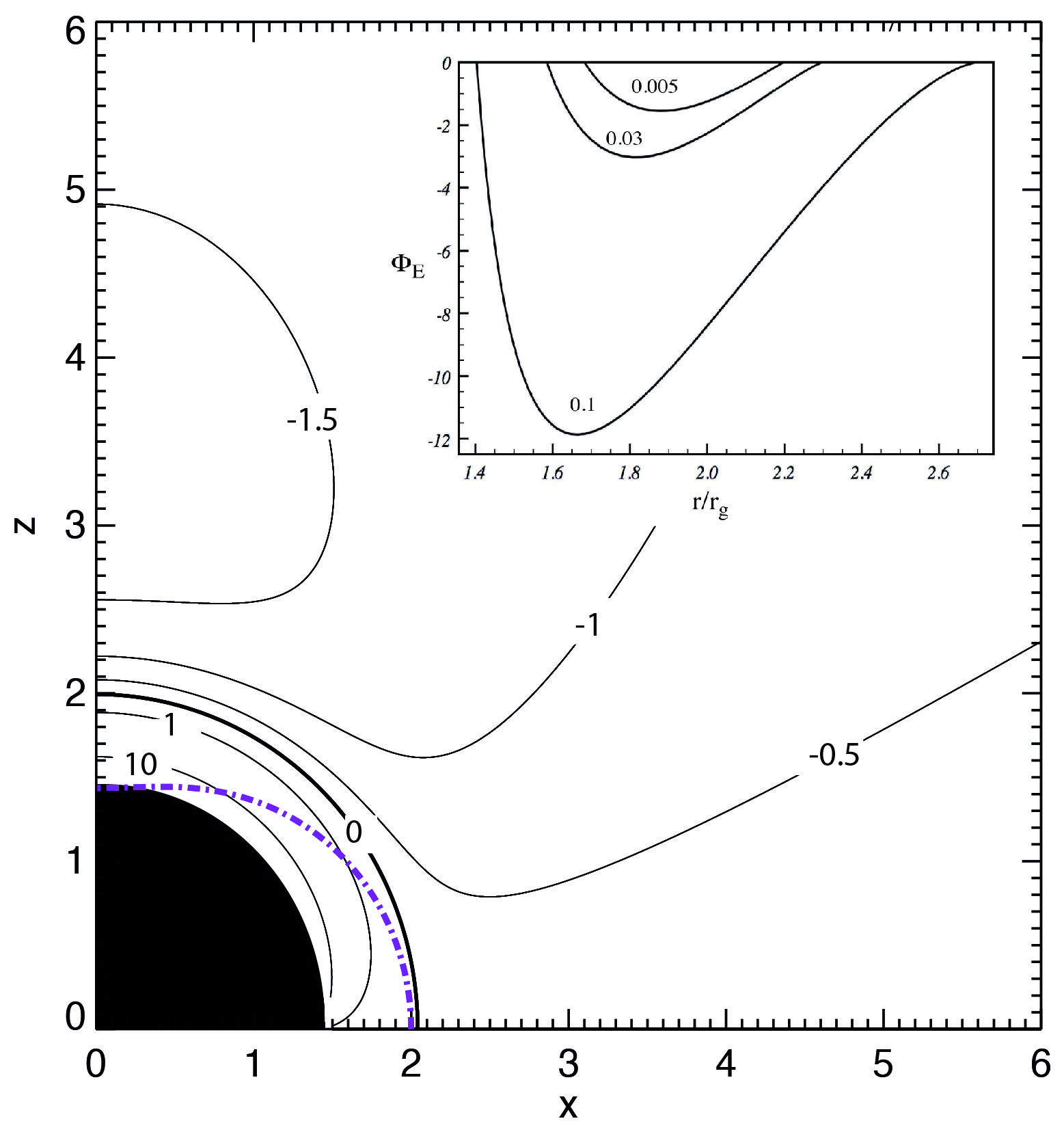} 
\caption{Contours of $\rho_{GJ}$ for a split monopole geometry with $a/M=0.9$, $\Omega=0.5\omega_H$. The numbers that 
label the curves are values of $\rho_{GJ}(r,\theta)$, normalized by $a B_H /2\pi$, where $B_H=\Psi_H/\sqrt{A_H}$ is the strength 
of the magnetic field on the horizon. The thick solid line corresponds to the null surface on which $\rho_{GJ}=0$. The black circle 
delineates the interior of the black hole, and the purple dashed-dotted line marks the static surface. 
The inset shows solutions for the electric flux, $\Phi_E=\sqrt{A}E_r^\prime$, computed from Eq. (\ref{eq:1D_Poiss_GR}) with 
$\rho_e<<\rho_{GJ}$ along a magnetic surface inclined at $\theta=30^\circ$, in a magnetosphere of a supermassive black hole of 
mass $M=10^9 M_\odot$, for $B_H=10^{4}$ G.}   
\label{fig:ex1}
\end{figure}

Stationary models tacitly assume that a gap forms in a localized region of the magnetosphere outside which the
ideal MHD condition (that is, ${\bf E}\cdot{\bf B} = 0$) prevails.   
To gain insight it is instructive to derive the gap electrostatic equation in flat spacetime first.  The generalization to 
Kerr spacetime then readily follows.
In general, the angular velocity of magnetic field lines,  ${\bf \Omega}=\Omega\hat{z}$, is conserved along magnetic 
surfaces only in regions where the ideal MHD condition is satisfied.  However, under the assumption that the gap 
constitutes a small disturbance in the global magnetosphere, the variation of ${\bf \Omega}$ across the gap (due to 
the finite potential drop) can be ignored. It is then convenient to transform to a rotating coordinate system,
$t^\prime=t, r^\prime=r, \theta^\prime=\theta$ and $\varphi^\prime=\varphi-\Omega\, t$, here in spherical coordinates,
in which the electric and magnetic fields are given by $ {\bf B}^\prime={\bf B}$, ${\bf E}^\prime={\bf E}+{\bf v}\times {\bf B}$ 
in terms of their components in the non-rotating frame (using geometric units, $c=1$), where ${\bf v}={\bf \Omega}
\times{\bf r}$ is the tangential velocity of the magnetic flux tube. 
It  can be readily shown that in the rotating frame Gauss's law takes the form
\begin{equation}
\nabla\cdot{\bf E}^\prime=4\pi(\rho_e-\rho_{GJ}),
\label{eq:Poiss_flat}
\end{equation}
where $\rho_e$ is the charge density and $\rho_{GJ}= -\nabla\cdot({\bf v}\times{\bf B})/4\pi$ denotes the Goldreich-Julian 
(GJ) density. In dipolar and split monopole geometries this simplifies to the well known result $\rho_{GJ} = -{\bf \Omega}
\cdot{\bf B}/2\pi$. Equation (\ref{eq:Poiss_flat}) indicates that a charge density $\rho_e=\rho_{GJ}$ is needed to screen 
out a parallel $E_{||}$.
The generalization to  Kerr geometry is straightforward. In Boyer-Lindquist coordinates, $(t,r,\theta,\varphi)$, the electric 
field component in the rotating coordinate frame is $F^\prime_{\mu t}=F_{\mu t}+\Omega F_{\mu\varphi}$, where 
$F_{\mu\nu}$ denotes the electromagnetic tensor, and we henceforth adopt the metric signature $(-,+,+,+)$.
The GR equations can be recast in a form similar to that in flat spacetime upon using quantities measured in the local
ZAMO (zero angular momentum observer) frame. For illustration, we adopt a split monopole geometry, defined by the 
potential $A_\varphi = \Psi_H(1-\cos\theta)$, with $\Psi_H$ being the magnetic flux on the horizon.
The non-corotating electric field, measured by a ZAMO, is then given by $E^\prime_r=\sqrt{A} F^\prime_{rt}/\Sigma$ in 
terms of the metric functions (in geometric units) $\Sigma=r^2 + a^2\cos^2\theta$ and $A= (r^2+a^2)^2- a^2\Delta 
\sin^2\theta$, with $\Delta=r^2+a^2-2M r$, $M$ and $a$ being the black hole mass and specific angular momentum (spin parameter), 
respectively. The general relativistic Gauss's law reads:
\begin{equation}
\partial_r (\sqrt{A} E_r^\prime) = 4\pi \Sigma (\rho_e-\rho_{GJ}),\label{eq:1D_Poiss_GR}
\end{equation}
where 
\begin{equation}
\rho_{GJ}=\frac{\Psi_H}{4\pi \sqrt{-g}}\partial_\theta \left[\frac{\sin^2\theta}{\alpha^2}(\omega-\Omega)\right]
\label{eq:rho_GJ_1D_split}
\end{equation}
is the general relativistic (GR) version of the GJ density, $\alpha=(\Sigma \Delta/A)^{1/2}$ is the lapse function and $\omega=
2aMr/A$ is the ZAMO angular velocity. In flat spacetime $\sqrt{A}=\Sigma=r^2$, and Eq. (\ref{eq:1D_Poiss_GR}) reduces to
Eq. (\ref{eq:Poiss_flat}) in spherical coordinates. The charge density is related to the proper densities, $n_\pm$, and the time 
component of the 4-velocities, $u^0_\pm$, of the e$^\pm$ pairs through $\rho_e =e(n_+u_+^0-n_-u_-^0)$.
Equation (\ref{eq:1D_Poiss_GR}) is subject to the condition ${\bf E}^\prime=0$ at the inner and outer gap boundaries, which 
follows from the assertion that the flow outside the gap is in a force-free state.  This implies that $E^\prime_r$ does not change 
sign, and that $|E_r^\prime|$ has a maximum inside the gap (see inset in Fig. \ref{fig:ex1}). An immediate consequence is that 
$\rho_{GJ}$ must change sign inside the gap \citep{Hirotani2016,Levinson2017}. Thus, the gap should form around the null 
surface on which $\rho_{GJ}$ vanishes. Such a null surface is a distinctive feature of a Kerr black hole that results from the frame 
dragging effect (see Fig. \ref{fig:ex1})\footnote{A null surface exists in a pulsar outer gap for other reasons}.
%
Now, from the above it is seen that at the null surface the gap electric field scales as $|E^\prime_{r}|\propto h^2$ with the gap 
width $h$, and the potential as $V\propto h^3$. The maximum electric current flowing through the gap is limited by 
$j^r=|\rho_{GJ}(r_1)| c \sim |d\rho/dr|_{0}\,h/2$, where $r_1$ is the outer gap boundary and the subscript $0$ designates values 
at the null surface. Thus, the maximum power that can be tapped, $L_{\rm gap}\simeq 2\pi r_H^2 j V$, and, hence, the maximum 
gamma-ray luminosity that can be emitted by the gap, scale as $h^4$.  A more precise expression for $L_{\rm gap}$ is derived in 
Ref.~\cite{Hirotani2016}, see also \citep{Greg2018}. Since the gap closure condition restricts the multiplicity inside the gap to 
unity, the gap width $h$ increases with decreasing pair production opacity. The salient lesson is that in steady gap models the 
output power of the gap increases steeply with decreasing disk luminosity. Detectability of gap emission favours low luminosity 
sources. We shall see shortly that this is not true in case of intermittent gaps.
A quantitative treatment of gap emission requires inclusion of plasma dynamics, radiation back-reaction, Compton scattering and 
pair production. Details are given in Refs.~\cite{Hirotani2016,Hirotani2016b,Hirotani2017,Levinson2017}. It is found that the 
characteristic spectrum produced in the gap consists of two components; curvature emission that peaks at sub-TeV energies, and  
inverse Compton emission that peaks at $10$ to $100$ TeV, depending on the black hole mass. Refs.~\cite{Hirotani2016,Hirotani2016b,
Hirotani2017,Lin2017} predict that gap emission from stellar and supermassive black holes should be detected by upcoming experiments.

\subsubsection{Time-dependent models}
\begin{figure*}[h]
\centering
\includegraphics[width=8cm]{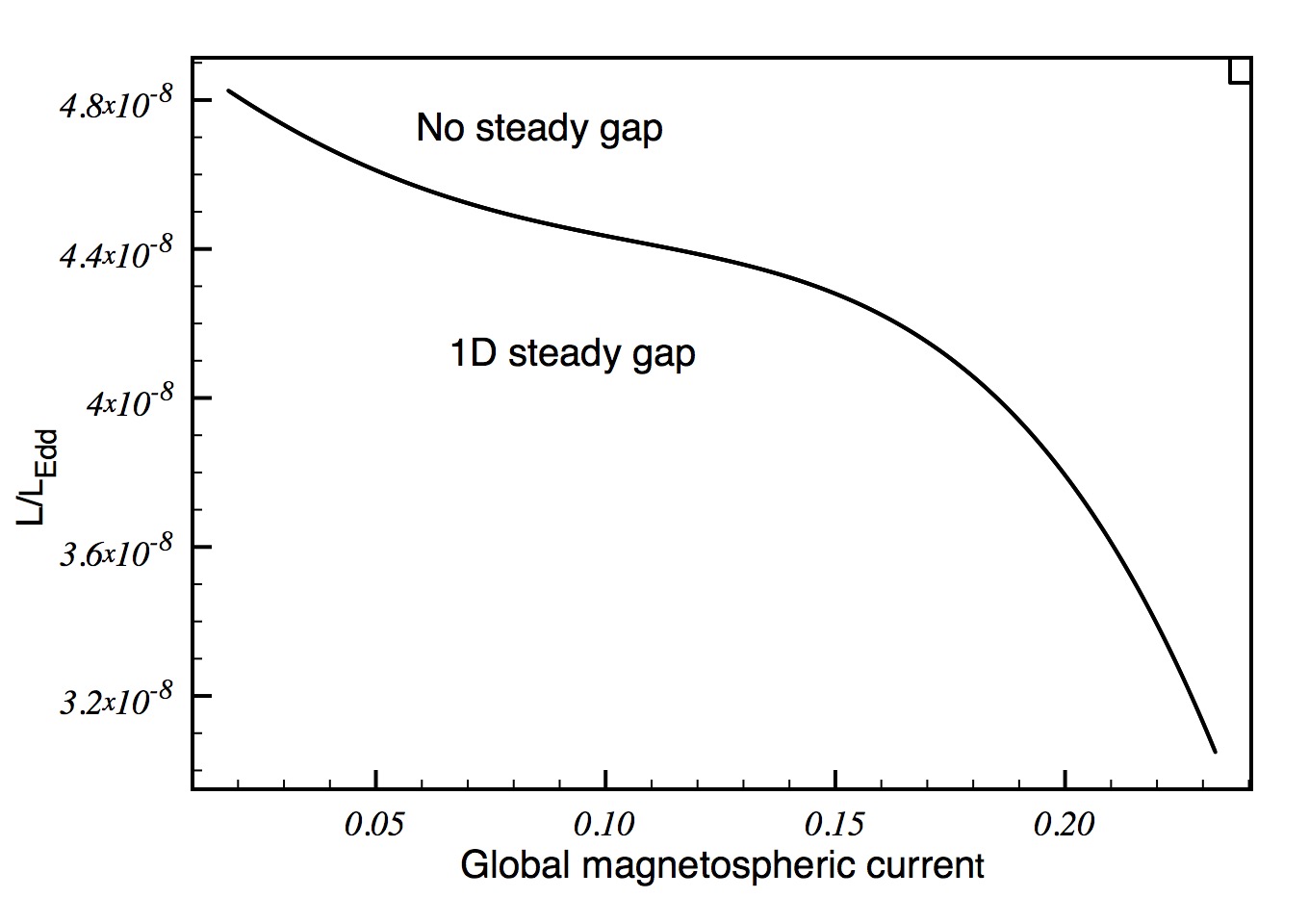} 
\caption{Maximum Eddington ratio below which local steady gap solutions exist, versus normalized magnetospheric current, 
$2\pi J_0/(\Sigma_{null}\Omega B_H\cos\theta)$, where $\Sigma_{null}$ is the value of $\Sigma$ on the null surface, for a 
supermassive black hole of mass $M_{BH}=10^9 M_\odot$, magnetic field $B=10^{4}$ G, a radiation source of size $R=30 r_g$ 
and a power law spectrum, $I_\nu\propto\nu^{-2}$, with a low energy cutoff at $\nu_{min}=10^{12}$ Hz, and inclination angle 
$\theta=30^\circ$. (See Ref.~ \cite{Levinson2017} for further details).
}   
\label{fig:stead-cond}
\end{figure*}

\begin{figure*}[t]
\centering
\includegraphics[width=9cm]{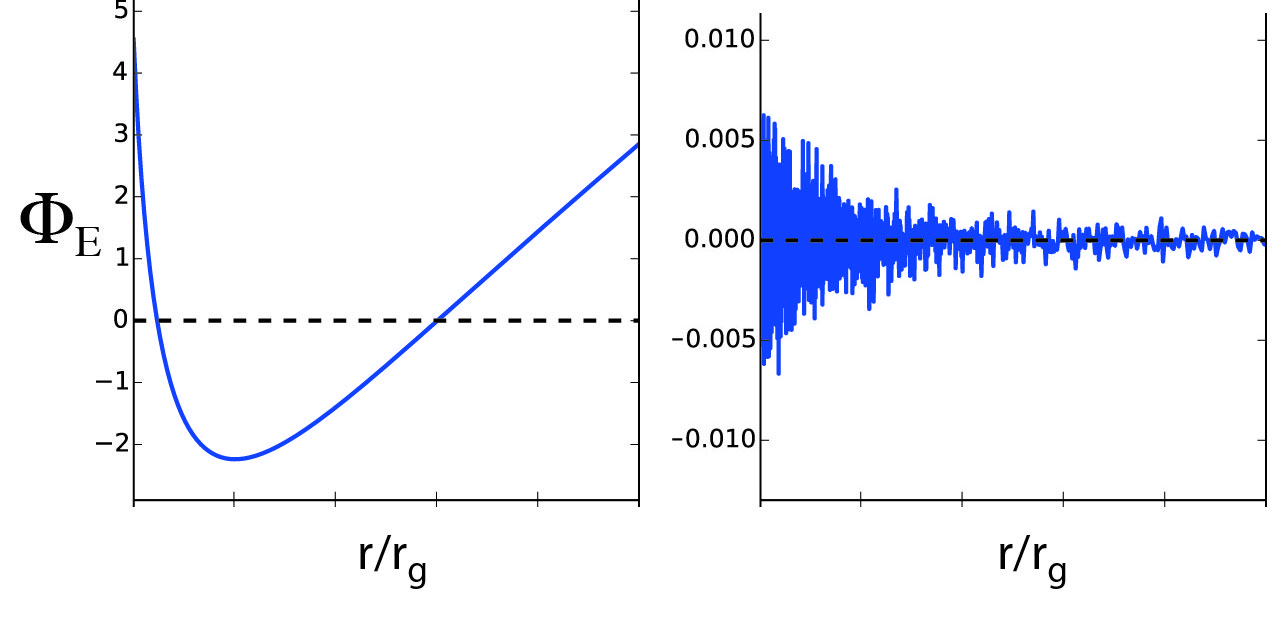} 
\caption{Evolution of the gap electric field in a magnetosphere of a supermassive black hole with the same parameters as in 
Fig.~\ref{fig:stead-cond}:  The left panel shows the electric flux $\Phi_E=\sqrt{A}E_r^\prime$ at the initial time $t=0$.
The right panel shows the electric flux at time $t=20 r_g/c$, roughly the onset of the quasi-steady oscillations.  Note the 
change of scale on the vertical axis between the two panels.  The amplitude of $E_r^\prime$ during the quasi-steady 
oscillations is about $10^{-3}$ of its initial magnitude at the null surface. The fiducial opacity in this case is $\tau_0=10$.}
\label{fig:E_evolution}
\end{figure*}

\begin{figure*}[h]
\centering
 \includegraphics[width=14cm]{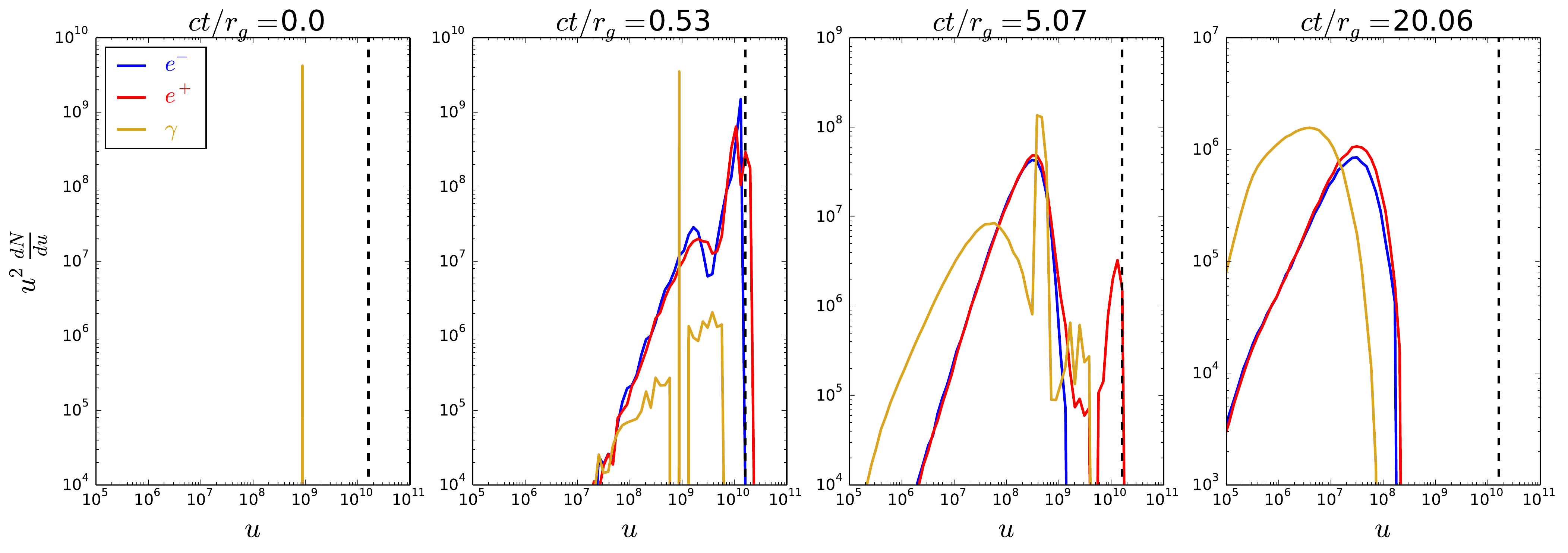} 
\caption{Evolution of the electron (blue), positron (red) and photon (yellow) spectra for the case shown in Fig. \ref{fig:E_evolution}. 
The photon energy is given in $m_ec^2$ units.  The peak of the quasi-steady gamma-ray spectrum in this example (rightmost panel) 
is at $\sim1$ TeV.}   
\label{fig:spectra}
\end{figure*}

The main difficulty with steady gap solutions for the null surface is that they exist only under highly restricted conditions, that may 
not apply to most objects \cite{Levinson2017}.  The reason is that the gap closure condition (i.e., unit multiplicity) imposes a relation 
between the global magnetospheric current (through the gap width) and the pair production opacity, which in reality are two 
independent and unrelated quantities.  An exemplary illustration for the existence regime of steady gap solutions is displayed in Fig. 
\ref{fig:stead-cond}, for a supermassive black hole of mass $m=10^9$ and a power-law disk emission spectrum. Similar results are 
obtained for stellar mass black holes. This example indicates that in practice, steady gaps can form only in sources with extremely 
low Eddington ratios. Additional difficulty stems from an inconsistency between the flow direction at the outer gap boundary and the 
ideal MHD flow below the stagnation surface \cite{Globus2014,Levinson2017}.  
Finally, the stability of steady gaps is questionable. The main conclusion is that  sparking of starved magnetospheric regions is likely 
to be inherently intermittent.
%
Attempts to construct 1D time-dependent models have been reported recently \cite{Levinson2018,Chen2018}.  
The analysis described in Ref.~\cite{Levinson2018} was performed using a newly developed, fully GR (in Kerr geometry) 
particle-in-cell code that implements Monte-Carlo methods to compute the interaction of pairs and gamma-rays with the soft 
photons emitted by the accretion flow.  As in the steady models mentioned above, the gap is assumed to be a small disturbance
that does not affect the global magnetospheric structure. The evolution of the electric field is governed by the equation 
\begin{equation}
\partial_t(\sqrt{A} E^\prime_r) =- 4\pi(\Sigma j^r-J_0),\label{dEdt}
\end{equation}
that generalizes the flat spacetime model derived in Ref.~\cite{Levinson2005}.  Here $j^r =e(n_+u_+^r-n_-u_-^r)$ is the radial 
electric current density, expressed in terms of the proper electron ($n_-$) and positron ($n_+$) densities and their 4-velocities 
$u_-$ and $u_+$, and $J_0$ represents the global magnetospheric current, which is an input parameter of the model in addition 
to $\Omega$.  The initial condition, $E^\prime_r(t=0,r)$, is obtained upon solving Eq. (\ref{eq:1D_Poiss_GR}) at the beginning 
of each run. For details the reader is referred to Ref.~\cite{Levinson2018}.
The analysis indicates that when the Thomson length for collision with disk photons becomes smaller than the gap width, i.e. 
$\tau_0>1$, screening 
of the gap occurs, following a prompt discharge phase that exhausts the initial energy stored in the gap, through low amplitude, 
rapid plasma oscillations that produce self-sustained pair cascades, with quasi-stationary pair and gamma-ray spectra. An example
is exhibited in Figs. \ref{fig:E_evolution} and \ref{fig:spectra}. The gamma-ray spectrum emitted from the gap peaks in the TeV 
band (Fig. \ref{fig:spectra}), with a total luminosity that constitutes a fraction of about $10^{-5}$ of the corresponding Blandford-Znajek 
power. This seems different than the spectrum predicted by steady gap models.  As those simulations are demanding, only a small 
range of parameters was explored in \cite{Levinson2018}. Future studies should investigate how the emission properties depend 
on the black hole mass, the target radiation spectrum and the magnetic field strength.
%
%
A caveat of the model described above is that it does not account for the effect of the ideal MHD fields on the dynamics of the plasma.   
An attempt to include such additional forces has been made in Ref.~\cite{Chen2018}, where it has been argued that the gap dynamics 
is cyclic in nature rather than quasi-steady. However, the analysis in this work is based on a mirror gap model in flat spacetime, that 
inherently assumes the presence of a current sheet (or current side flow) at the stagnation surface (by reversal of the toroidal magnetic 
field there).  Whether this model is a reasonable representation of a starved Kerr black hole magnetosphere remains to be demonstrated.   
A cyclic behaviour is naively anticipated in a global model that links the gap and the magnetosphere together \cite{Levinson2017}.  
How this would affect the spectrum and light curves of the HE gap emission is yet an open issue that can only be studied using 2D 
simulations.

\subsection{Variable VHE and inner jet models}
\begin{figure}
\includegraphics[angle=0, width=4.95cm]{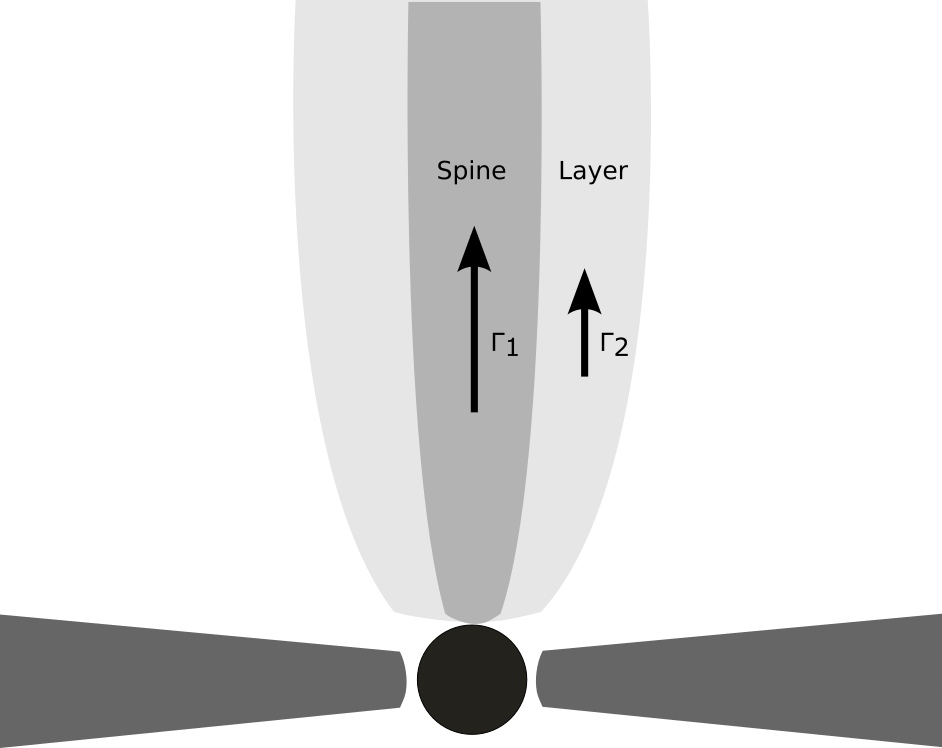}
\hspace{0.1cm}
\includegraphics[angle=0, width=4.95cm]{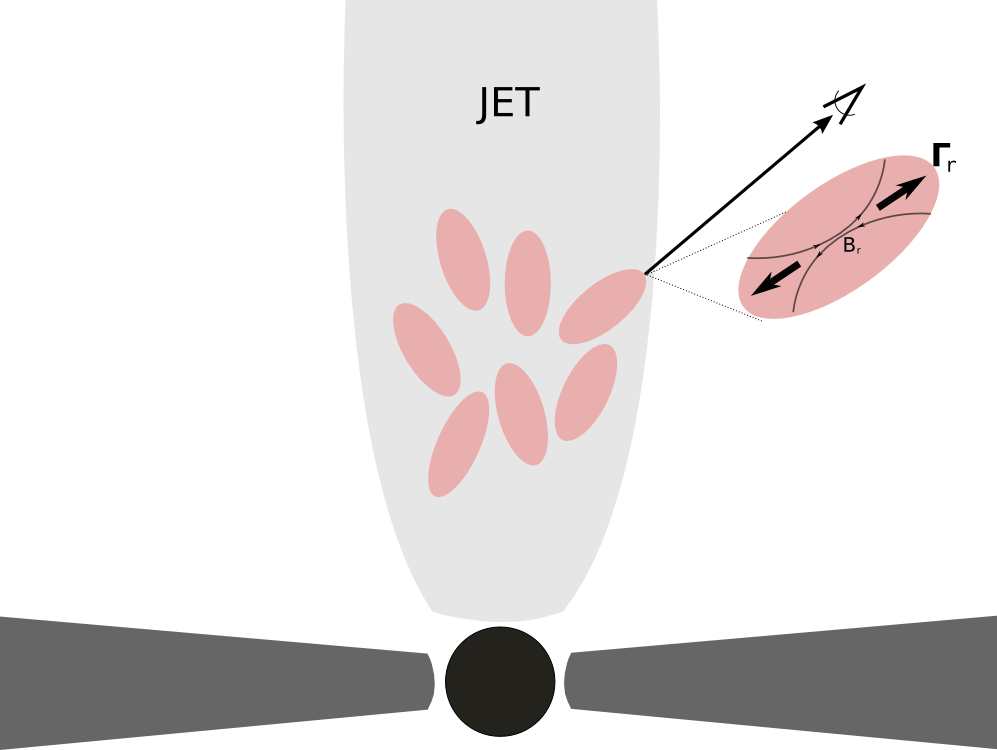}
\hspace{0.1cm}
\includegraphics[angle=0, width=4.95cm]{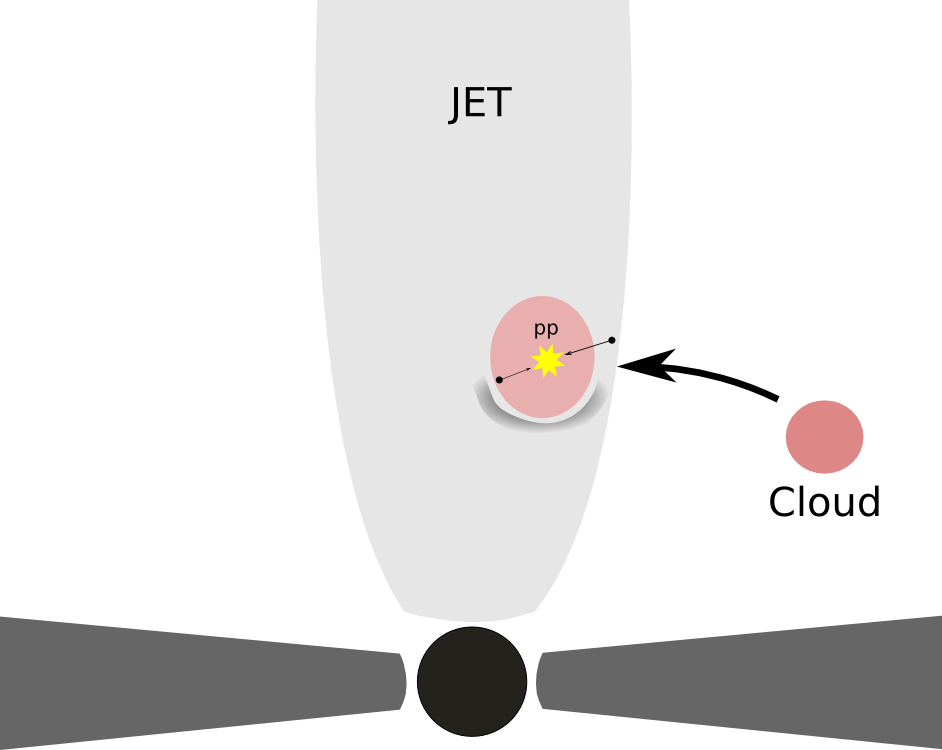}
\caption{Possible scenarios for the origin of rapidly variable VHE emission on scales of the inner jet, i.e. beyond black hole 
magnetospheric emission. Left: Illustration of a stratified jet composed of a fast spine - slow layer \citep{Tavecchio2008}. 
Radiative interplay (external IC) can increase the $\gamma$-ray luminosity of each component. Middle: Illustration of 
a jets-in-jet model \cite{Giannios2010}, where a variety of 'mini-jet' features (plasmoids) are triggered by magnetic 
reconnection events within the main jet flow. This could lead to an additional velocity component relative to the main flow 
($\Gamma_r$) and allow a favorable orientation with respect to the observer. Right: Illustration of a hadronic model, 
where interactions of the jet with a massive obstacle (star or cloud) facilitates shock-acceleration and introduces a 
sufficient target density to allow for efficient $pp$-collisions \cite{Barkov2012}.}\label{models}
\end{figure}

Rapid variability on timescales of days and below implies a compact emitting region, and is often taken to indicate that this
emitting region is located at sub-parsec jet scales ($<10^4\, r_g$) and below. Promising jet-related proposals in this context 
(beyond black hole gaps) include the following (cf. Fig.~\ref{models}):

\subsubsection{Spine-shear scenarios}\label{spine-shear}
The complex interplay of black hole- and disk-driven outflows along with environmental interaction (e.g., entrainment) is likely to give 
rise to a non-uniform flow topology where different jet layers possess different bulk flow speeds. In its simplest realization a 
fast ($\Gamma_1\gg 1$) spine is taken to be surrounded by a slower-moving ($\Gamma_2<\Gamma_1$) sheath or layer 
\citep[e.g.,][]{Ghisellini2005}. This is essentially a two-zone model, yet with parameter-space constrained by the requirement
of an efficient radiative interplay. Depending on the viewing angle we may see different parts of the jet (spine or sheath), 
resulting in different emission characteristics for misaligned AGN when compared with those of blazars. In particular, given 
their larger inclination angles, the VHE emission in RGs would usually be dominated by the stronger (!) Doppler-boosted 
emission from the layer. VHE variability then essentially imposes constraints on the size of the layer. Spine-sheath models 
can in principle accommodate fast variability and facilitate an increased gamma-ray luminosity as each component sees the 
(external) radiation of the other amplified by the relative motion between them ($\Gamma_{\rm rel}=\Gamma_1\Gamma_2 
[1-\beta_1\beta_2]$), thereby enhancing the IC contribution of both its spine and its layer \citep{Ghisellini2005,Tavecchio2008,
Tavecchio2014}. In order for this interplay to work efficiently however, the emitting zones need to be (quasi) co-spatial. This 
in turn leads to transparency issues as long as one aims at a simultaneous reproduction of the infrared-optical part of the 
SED, since the large gamma-ray opacity associated with the intense (weakly Doppler-boosted) radiation field of the spine 
usually results in considerable $\gamma\gamma$-absorption and a steep slope at TeV energies. Hard TeV spectra thus 
cannot simply be accounted for with such models \citep[e.g.,][]{MAGIC2018}. Nevertheless, an internal velocity stratification
(shear) bears the potential to solve several issues regarding the unification of BL Lacs and RGs \citep{Chiaberge2000,
Meyer2011,Sbarrato2014}. The noted emission models represent a simplified first approach, that focus on the radiative 
interplay only and do not yet take any acceleration effects \cite[e.g.,][]{Rieger2004} nor time-dependencies into account. 
Decoupling of the infrared and VHE emission parts would alleviate some tensions, though it would leave a larger part of 
the parameter space unconstrained. Similar to other two-zone approaches, the non-thermal emission can in many cases
be reproduced under equipartition conditions which may be counted in its favour \citep{Tavecchio2016}. Time-dependent 
extensions are certainly needed to further assess its potential in the context of RG modelling. 

\subsubsection{Reconnection - mini-jets and plasmoids}
Relativistic jets are generally expected to be initially magnetically dominated (magnetization $\sigma_m\gg 1$), and this has 
motivated studies where the field energy is released via collisionless magnetic reconnection allowing for non-thermal particle 
acceleration and subsequent high energy gamma-ray (IC) production \citep[e.g.][]{Giannios2009,Giannios2010,Nalewajko2011,
Cerutti2012,Giannios2013,Kadowaki2015,Sironi2015,Werner2016}. An open question at present concerns whether formation 
of current sheets on sufficiently small scales can result from current driven instabilities induced during the propagation of the 
jet \cite[e.g.,][]{mizuno2012,oneil2012,guan2014}, or may inherently form during injection of the jet, e.g., due to advection of 
asymmetric magnetic field or magnetic loops by the disk \cite[e.g.,][]{parfrey2015,levinson2016}. The observed collimation 
profile of M87 seems to indicate that in this source the jet is kink stable \cite{Globus2016}. Upcoming EHT observations on 
horizon scale  may shed more light on this issue.\\
Reconnection-related models of the type proposed in \citep{Giannios2010} assume that relativistic (Petschek-type) reconnection 
occurs in the jets of RGs (taking them to be electron-proton dominated with $\sigma_m\sim 100$), leading to efficient electron 
acceleration and the generation of variable VHE gamma-rays via  inverse Compton (SSC and/or EC) scattering \cite[e.g.,][]{Cui2012}. 
Moreover, efficient reconnection allows for an additional relativistic velocity component of the ejected plasma ($\Gamma_r \simeq 
\sqrt{\sigma_m}$) relative to the mean bulk flow of the jet, and directed at some angle to it. Given multiple, localised reconnection 
sites ("mini-jets") within the jet, strong differential Doppler boosting effects ($D\gg1$) could become possible even for RG sources 
whose main jet direction are substantially misaligned. This could then provide a simple explanation for ultra-fast variability in 
misaligned AGN. Caveats concern whether such high magnetisations should indeed be expected for electron-proton (disk-driven) 
jets, and whether the impact of a magnetic guide field, that would lead to weak dissipation only \citep{Lyubarski2005}, can be 
neglected on the anticipated scale of VHE gamma-ray production. Nevertheless, reconnection is particularly interesting as it 
could facilitate non-thermal particle energization in those magnetized part of the jets where diffusive shock acceleration is 
inefficient \citep[e.g.,][]{Lemoine2017}, and ensure a rough equipartition between magnetic fields and radiating particles in the 
reconnection downstream (emitting) region \citep{Sironi2015}. In the absence of a guide field, relativistic reconnection results 
in a flat (hard) power-law tail $n(\gamma)\sim \gamma^{-\alpha}$ with slope $\alpha<2$ for $\sigma_m>10$, most probably 
approaching $\alpha \sim1$ in the extreme relativistic case \citep{Zenitani2001,Werner2016,guo2015}. For large Lundquist numbers 
the reconnection layer becomes unstable to resistive (tearing) instabilities, causing its fragmentation into local magnetic islands or 
plasmoids. More recent studies have thus concentrated on characterizing the (fractal-like nature of) plasmoid formation in 
relativistic pair plasmas \cite{Sironi2016,Petropoulou2018}.

\subsubsection{Jet-star/cloud interactions}
The ubiquity of stars and gas in the central region of the AGN host galaxy can lead to frequent jet-star/cloud interactions, facilitate jet 
entrainment and mass loading by stellar winds \cite[e.g.,][]{Perucho2017}, and give rise to knotty X-ray structures as e.g. seen in the 
large-scale jet of Cen~A \citep{Wykes2015}. Over the last few years a variety of studies have explored its impact on the 
gamma-ray emission characteristics in AGN, both for its steady and transient/flaring states \citep[e.g.][]{Barkov2010,Bosch-Ramon2012,
Araudo2013,Khangulyan2013,Bednarek2015,Bosch-Ramon2015,Cita2016,Zacharias2017,Vieyro2017}. In the RG context, one interesting 
application concerns the possible contribution of inelastic proton-proton (pp) collisions in the generation of variable VHE emission. While 
AGN jets are commonly considered not to carry enough target matter ($n_p$) to allow for efficient pp-collisions (given its long timescale 
$t_{pp} \simeq 10^{15}/n_p$ sec), interactions of a red giant star or a massive, dense gas cloud (size $r_o$) with the base of the jet (radius 
$r_j$) could occasionally introduce a high amount of matter, trigger shock acceleration and potentially drive rapid VHE activity 
\cite{Barkov2010,Barkov2012}. Model calculations in the case of M87 \citep{Barkov2012b} suggest that such a scenario could account 
for the observed VHE characteristics (including day-scale variability) if the jet would be powerful enough and a sufficiently large fraction 
($\propto r_o^2/r_j^2$) of it could be channeled into VHE $\gamma$-ray production. Given the observed large opening angle and 
transversal dimension $r$ of the milli-arcsec radio jet in M87, the latter is not obvious. In fact, simple models of this type often need jet 
powers in excess of current estimates. In principle, however, these power constraints could be somewhat relaxed if the jets possess a 
spine-shear-type configuration with most of the energy flux concentrated into a narrow core. This would then bear some similarities with 
the set-up discussed in Sec.~\ref{spine-shear}. In addition, the effective size $r_0$ (if related to shocks at quite some distance from the 
star) could well be larger than the initial size of the obstacle \citep[e.g.,][]{Bosch-Ramon2015,Bednarek2015}. While the availability of 
suitable stellar orbits on sub-parsec scales limits the possible recurrence (frequency) of short, star-driven VHE flaring events, multiple 
collisions along the jet seem unavoidable. Depending on conditions in the host galaxy, this could also result in a detectable, steady 
$\gamma$-ray contribution at VHE energies; a recent calculation of the cumulative emission from multiple jet - stellar wind interactions 
in M87, however, suggests this to be too low to account for its overall VHE flux levels \citep{Vieyro2017}.

\subsection{Steady VHE and extended jet models}
\begin{figure*}[h]
\centering
 \includegraphics[width=10cm]{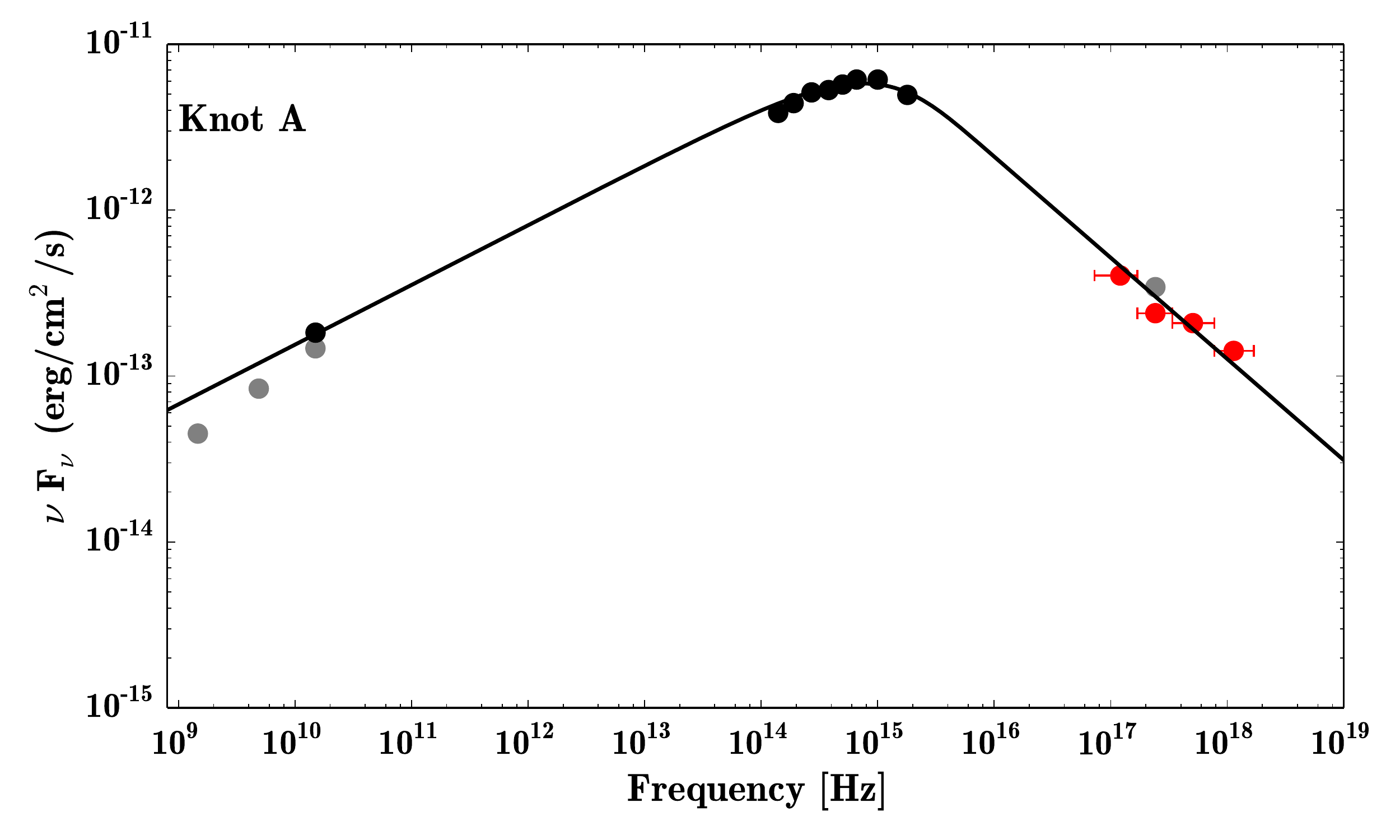} 
\caption{Representative multi-wavelength SED for the outer knot A in the kpc-scale jet of M87 including X-ray emission (red) 
based on 1.5 Msec of Chandra data. 
The emission is thought to be synchrotron in origin, indicating the presence of electrons up to multi-TeV energies within the jet. 
Inverse Compton up-scattering off various soft photon fields (starlight, dust, EBL, CMB) would result in a steady VHE contribution. 
From Ref.~\citep{Sun2018}.}
\label{M87_knot}
\end{figure*}

The detection of extended X-ray emission from the large-scale jets in AGN by Chandra \footnote{https://hea-www.harvard.edu/XJET/} 
has raised the possibility that these jets are also steady sources of VHE $\gamma$-rays. Electron synchrotron radiation is by now the 
favoured interpretation for the X-ray emission \citep[e.g.,][]{Harris2006,Georganopoulos2016}, indicating the presence of highly energetic 
electrons with Lorentz factors up to $\gamma \sim 10^8 (100\,\mu\mathrm{G}/B)^{1/2} D^{-1/2}$, where $D$ is the Doppler factor and 
$B$ the large-scale magnetic field strength.
While diffusive shock acceleration in the jet could in principle facilitate such energies, localized (shock-type) acceleration at knots 
is usually not sufficient given the fact that there is little evidence for e.g. the inter-knot regions to have significantly steeper spectra 
than the adjacent knots as one would expect in the case of synchrotron cooling. This may point to the operation of a continued or
distributed acceleration mechanisms such a stochastic or shear particle acceleration \citep[e.g.,][]{Liu2017}.\\
Figure~\ref{M87_knot} shows a SED result of a recent spectral analysis in M87 \citep{Sun2018}. The difference of the radio and 
X-ray spectral indices supports a synchrotron as opposed to an IC origin. 
%
The multi-TeV electrons in the jets will then also up-scatter the host starlight, dust, EBL or CMB photons to $\gamma$-ray energies, 
resulting in a weak and steady VHE contribution that may become detectable in nearby RG. IC up-scattering of CMB ($u_{CMB}
=4.2\times10^{-13}$ erg cm$^{-3}$; $\nu_p\simeq 1.6\times 10^{11}$ Hz), for example, leads to a non-reducible multi-TeV 
contribution at flux levels of $f_{IC} \sim (u_{IC}/u_B) f_x \sim 10^{-3} f_x$. For M87 however, the total X-ray flux of the knots is 
of the order $f_x \sim 10^{-12}$ erg cm$^{-2}$s$^{-1}$ only, implying an IC-CMB contribution at flux levels well below current 
detections. Up-scattering of dust or starlight photons on the other hand, is expected to yield a much higher VHE contribution that 
could be probed with CTA \citep{Hardcastle2011}. Given the (current) absence of VHE variability in Cen A, an extended (leptonic)
origin of its VHE emission certainly cannot be excluded. In addition, the cumulative IC emission from multiple jet-star collisions in 
the kpc-scale jet \cite{Bednarek2015} might further contribute. Evidence for a possible VHE extension has in fact been recently
reported for Cen~A \citep{Sanchez2018}. We note that an experimental verification of extended VHE emission would support the 
notion that the large-scale jets in AGN could make a relevant contribution to the TeV background when compared to the 
highly-boosted VHE emission from blazar cores \citep{Georganopoulos2016}.

\section{Conclusions}
The experimental progress over the last decade has led to the discovery of radio galaxies (RGs) at $\gamma$-ray energies, revealing 
exceptional features such as spectral hardening or ultra-fast variability. With their jets misaligned and related Doppler boosting effects
only modest, RGs are offering unique insights into physical mechanisms and environments (e.g., the plasma physics of jets or the black 
hole vicinity) that are otherwise difficult to access.\\ 
The unexpected spectral hardening at gamma-ray energies seen in Cen~A, for example, points to the emergence of a new physical 
component beyond the conventional SSC-type one, with current interpretations ranging from the smallest (sub-pc) to the largest (kpc) 
jet scales, up to extended dark matter scenarios. On the other hand, rapid VHE variability on timescales shorter or comparable to the 
light travel time across the horizon of the black hole, as e.g. seen in IC~310 and M87, provides evidence for a highly compact emitting 
zone, probably at the very origin of the jet itself (black hole gaps), or internally- (reconnection) or externally (star collision)-induced.\\
These and related experimental findings at gamma-ray energies have been vital in triggering important conceptual progress in black
hole - jet physics and generated a variety of promising research avenues. Along with further theoretical efforts, dedicated observational 
studies are currently needed to e.g. clarify source classification and to better resolve their timing characteristics. The upcoming CTA 
array \citep{CTA2017} will have the potential to probe deeper into the spectral and variability characteristics of RGs, and thereby allow
to fundamentally advance our understanding of the AGN phenomena in general.

\vspace{6pt} 
\funding{FMR acknowledges funding by a DFG Heisenberg Fellowship RI 1187/6-1. AL acknowledges support by 
the Israel Science Foundation (grant 1114/17).}

\acknowledgments{FMR acknowledges the support and hospitality of the MPIK Heidelberg.}


\end{document}